\documentclass[preprint]{aastex}
\usepackage{lscape}

\begin{document}

\title{Observational Constraints on the Formation and Evolution of 
Binary Stars}

\author{R. J. White\altaffilmark{1} and A. M. Ghez}

\affil{UCLA Division of Astronomy and Astrophysics, Los
Angeles, CA 90095-1562}

\altaffiltext{1}{Present Address: McDonald Observatory, R.L.M. Hall 15.308,
Austin, TX 78712-1083}

\begin{abstract}

We present a high spatial resolution multi-wavelength survey of 44 
young binary star systems in Taurus-Auriga with separations of 10 - 1000
AU.  These observations, which were obtained using the Hubble Space
Telescope and the NASA Infrared Telescope Facility, quadruple the number of
close ($<$ 100 AU) binary stars with spatially resolved measurements from
0.3 to 2.2 $\mu m$ and are the first 3.6 $\mu m$ measurements for the
majority of the companion stars in the sample.

Masses and ages are estimated for the components observed at optical
wavelengths.  The relative ages of binary star components are more similar 
than the relative ages of randomly paired single stars within the
same star forming region.  This is the first statistically significant
evidence for coeval formation.  Only one of the companion masses is
substellar, from which we conclude that the apparent overabundance of T
Tauri star companions relative to main-sequence star companions is not due
to a wealth of substellar secondaries that would have been missed in
main-sequence surveys.

The circumstellar environments of binary star systems are studied in this
work through three diagnostics - the infrared color $K-L$, the ultraviolet
excess $\Delta U$, and H$\alpha$ emission.  Several conclusions are drawn.
First, the mass accretion rates for primary stars are similar to single
stars, which suggests that companions as close as 10 AU have little effect
on the mass accretion rate.  Second, although most classical T Tauri star
binaries retain both a circumprimary and a circumsecondary disk, there are
several systems with only a circumprimary disk.  Systems with only a
circumsecondary disk are rare.  This suggests that circumprimary disks
survive longer than circumsecondary disks.  Third, primary stars accrete at
a higher rate, on average, than secondary stars.  This is most likely
because of their larger stellar mass, since the mass accretion rates for
both single and binary T Tauri stars exhibit a moderate mass dependence.
Fourth, approximately 10\% of T Tauri binary star components have very red
near-infrared colors ($K-L > 1.4$) and unusually high mass accretion rates.
This phenomenon does not appear to be restricted to binary systems,
however, since a comparable fraction of single T Tauri stars exhibit the
same properties.  These high accretion stars are probably not at an earlier
stage of evolution, as has been proposed.  Their semblance of younger
protostars at optical and infrared wavelengths is most likely because of
their similar high levels of accretion, which are above the norm for T
Tauri stars, and not because of similar ages.

The stellar and circumstellar properties are also used to indirectly trace
the evolution of circumbinary material.  In contrast to single T Tauri
stars, which have disk dissipation timescales comparable to their ages,
the disk dissipation timescales for binary T Tauri stars are roughly 1/10
of their ages.  Replenishment of the inner circumstellar disks may be
necessary to explain the continuing disk accretion in these systems.
The longer disk lifetimes of circumprimary disks, despite
their higher depletion rates, suggests that circumprimary disks are being
preferentially replenished, possibly from a circumbinary reservoir with low
angular momentum relative to the binary.   Further support for circumbinary
reservoirs comes from the observed correlated presence of circumprimary and
circumsecondary disks for binaries with separations of less than $\sim 200$
AU.  The presence of disks appears uncorrelated for wider binaries.
Additionally, binaries with separations of less than $\sim$ 100 AU exhibit
a higher fraction of high mass ratio (m$_\textrm{\scriptsize
s}$/m$_\textrm{\scriptsize p}$) pairs than wider binaries.  
These separation dependent properties can be explained if the components
are being replenished from a common circumbinary reservoir with low angular
momentum.  The components of the closest pairs are expected to be more
equally replenished than the widest pairs, which consequently sustains both
disks and drives their mass ratio toward unity.  Overall, the results of
this study corroborate previous work that suggests fragmentation is the
dominant binary star formation mechanism; disk instabilities and capture
seem unlikely.

\end{abstract}


\section{Introduction}

Surveys of young low mass stars in nearby star forming regions have
established that the majority are members of binary star systems
\citep[e.g.,][]{gnm93, leinert93, simon95}.  More than half of these
binary T Tauri stars have separations less than 100
AU, the characteristic size of a circumstellar disk
\citep[e.g.,][]{beckwith90}.  The ubiquity of these close
companions has raised two important and related questions concerning the
star formation process: ``How do binary stars form?'' and ``How does a
companion affect the distribution of circumstellar material?''.
Unfortunately, obtaining the answers to these questions has been inhibited
by the inherent difficulty of spatially resolving binaries with small
angular separations.  At the distance of nearby regions of star formation
(D $\sim$ 150 pc), the majority of these pairs have separations of
less than 1 arcsecond and are unresolvable with standard ground-based
observing
techniques.  Most T Tauri binary stars have only been spatially resolved at
the one wavelength, typically 2.2 $\mu$m, used in the high resolution lunar
occultation or speckle imaging surveys that revealed their multiplicity
\citep[e.g.,][]{gnm93, leinert93, simon95}.  Extracting the stellar and
circumstellar properties (e.g., stellar temperature and luminosity, mass
accretion rate) from this one resolved measurement is not possible since
young stars have substantial excess emission and line-of-sight extinction.
Determining these properties requires spatially resolved measurements over
a broad range of wavelengths.  A few studies have had some success
accomplishing this for wide binaries, or for small samples of close
binaries, and the combined results of these studies have been useful in
understanding how binary stars form \citep{ck79, hss94, koresko95, bz97,
gws97, wlk01}.  Although the samples are small, the mass ratio and
secondary mass distributions of T Tauri binary stars, as well as their
comparable ages, tentatively suggest that core fragmentation
\citep[e.g.,][]{boss88}is the dominant binary star formation process, while
capture and disk instability scenarios seem unlikely \citep{gws97}.

Current models of the binary star formation process offer a
framework with which we can begin to explore the distribution of
circumstellar material in binary star systems.  Numerical simulations
of circumstellar material within a binary suggest that both circumprimary
and circumsecondary disks are possible, as well as circumbinary structures
\citep{al94, bb97}.  The inner disks are expected to be
tidally truncated at roughly 1/3 of the semi-major axis of the binary
\citep{al94}.  Few observational constraints are available, however, to
check these predications.  Many spatially unresolved binary T Tauri stars
exhibit strong H$\alpha$ emission and photospheric excesses
suggesting the presence of an accreting inner circumstellar disk.  However,
it is in general not known whether this material is distributed in a
circumprimary disk, a circumsecondary disk, or both.  The presence of any
accreting disks in systems with separations less than 100 AU is somewhat
surprising given the small mass of these presumably truncated disks,
as inferred from millimeter measurements of the unresolved pairs
\citep[M$_\textrm{\scriptsize disk}$ $\lesssim 0.01$ M$_\odot$;][]{ob95,
dutrey96, jensen96}.  If the accretion rates for these close binaries are
similar to the rates for single T Tauri stars \citep[$\sim 10^{-8}$
M$_\odot$yr$^{-1}$;][]{gullbring98}, and if their ages are $\sim$ 1 Myrs,
many of these disks should have already dissipated.  This potential
timescale problem may be alleviated, however, if the inner circumstellar
disks are being replenished from a circumbinary reservoir \citep{ps97}.

Several spectroscopic studies have identified circumprimary disk and
circumsecondary disk accretion signatures for relatively wide T Tauri
binaries ($\gtrsim 100$ AU) resolvable with ground-based spectroscopy
\citep{bz97, ps97, mmd98, duchene99b, white99}.  This work showed that at
least for
widely separated pairs, both circumprimary and circumsecondary disks can
exist.  A variety of high resolution studies have been carried out in
order to resolve the properties of the closest and potentially most
interesting pairs with separations of less than 100 AU.  These studies
are typically conducted at near-infrared (NIR) wavelengths where high
resolution techniques such as speckle imaging or adaptive optics are more
realizable.  The majority of these studies, however, include only one
or a few multiple systems and typically focus on peculiar systems with
exceptionally NIR bright companions (e.g., T Tauri: Ghez et al. 1991; XZ Tau:
Haas et al. 1990) or with extended circumstellar emission (GG Tau:
Roddier et al. 1996; UY Aur: Close et al. 1998).  Only one extensive
NIR survey of close ($< 100$ AU) binaries has been carried out
\citep{wlk01}.  While the components of many of these systems show
signatures of retaining a circumstellar disk, details are difficult to
interpret since the underlying stellar emission is unknown; the stellar
properties are best extracted from optical measurements where the
photospheric emission dominates and where the spectral types and colors of
late-type stars are well established \citep[e.g.,][]{luhman99, white99}.  A
limited number of close binary T Tauri stars with signatures of
accretion have been imaged with high resolution optical techniques
(DD Tau: Bouvier et al. 1992; XZ Tau: Krist et al. 1997, 1999; FS
Tau: Krist et al. 1998; DF Tau, GG Tau, UZ Tau: Ghez et al. 1997).  No
complete optical survey of close ($<$ 100 AU) T Tauri binary stars has
yet been conducted.

In this paper, we present spatially resolved observations of 44 T Tauri
multiple star systems with projected separations of 10-1000 AU in the
nearby star forming region Taurus-Auriga \citep[D = 140 pc;][]{kenyon94}.
These 0.3 to 3.6 $\mu m$ observations represent the largest, most complete,
broad-wavelength survey of multiple T Tauri stars to date.  The source
list, observations, and analysis are described in \S 2.  The stellar and
circumstellar properties are derived in \S3 and then used in \S4 to
understand how circumstellar material is distributed in binary star
systems.  The results are also used to investigate the accretion rate's
dependence on the stellar mass, to determine the evolutionary state of
infrared-bright, high accretion stars, to further constrain how binary
stars form, and to confirm the over-abundance of T Tauri stellar companions
in Taurus-Auriga relative to main-sequence stars.

\section{Observations, Data Analysis, and Results}

\subsection{Sample Definition}

The observational objective of this program was to observe a
complete sample of multiple T Tauri star systems in the Taurus-Auriga
star forming region with separations ranging from 0\farcs07 - 7\farcs0
(10 - 1000 AU).  The lower
limit corresponds to the smallest separations resolvable with speckle
imaging on a 3-m telescope at $2\,\mu$m (i.e., $\lambda/2D$) and
the upper limit insures a high probability of physical association
\citep[$\sim$ 99\%,][]{rz93}.  In this survey, binaries with separations of
10 - 100 AU are referred to as \textit{close binaries} and those between
100 - 1000 AU are \textit{wide binaries}.  The initial target list, which
was constructed in 1995, consists of 50 pairs in 47 systems; systems in
this study are groups of stars with separations less than 10\farcs0.  
The majority of the binaries were identified from the high resolution
multiplicity surveys of \citet{gnm93}, \citet{leinert93}\footnote{The
binary stars HBC 351 and HBC 358 were not included because their low Li I
abundance \citep{walter88, martin94} suggests that they are not T Tauri
stars \citep{martin98}}, \citet{richichi94}, and \citet{simon95}.
Additional wide binaries were extracted from \citet{ck79}, \citet{hb88},
and \citet{hartmann91}.  Of these 50 pairs, 46 have been observed in this
study (DI Tau, HBC 412, Haro 6-10, and HV Tau were not observed) and 43
were resolved (V410 Tau A-B is unresolved and RW Aur Ba-Bb and FY Tau A-B
are probably not binary stars; see \S2.3.2).  Two additional companions
discovered
in these observations (V410 Tau C and FW Tau C) and supplemental
measurements from the literature (including HV Tau and Haro 6-10) increase
the total number of pairs resolved at multiple wavelengths to 47 (44
systems).  More specifically, 41 pairs (39 systems) have been spatially
resolved at both 2.2 and 3.6 $\mu m$ (\S 2.2) and 28 pairs (25 systems)
have been spatially resolved at optical wavelengths (\S 2.3; includes the T
Tau system, although T Tau B was not detected).  Table \ref{tab_sample}
provides a summary of the targets and observations included in this
analysis.

Both classical T Tauri star (CTTS) systems and weak-lined T Tauri star
(WTTS) systems are included in the larger sample of sources observed at
NIR wavelengths, whereas the smaller sample of sources observed at optical
wavelengths focuses primarily on the CTTS (Table \ref{tab_sample}).  CTTSs
are young stars that display signatures of active accretion such as strong
H$\alpha$ emission and large UV excesses.  In contrast, WTTSs are young
stars that are not thought to be accreting.  These stars exhibit only
'weak' H$\alpha$ emission and little if any UV excess, both of which can be
attributed entirely to an active chromosphere \citep[e.g.,][]{walter88}.
CTTS and WTTS systems are distinguished here based on the strength of
H$\alpha$ emission, usually in the unresolved spectra of the pair, and a
classical/weak dividing value set by \citet{martin98}.  A negative
equivalent width (EW) indicates emission.  T Tauri stars are considered
classical if their EW(H$\alpha$) $\le$ -5 \AA\, for K stars, $\le$ -10
\AA\, for M0-M2 stars and $\le$ -20 \AA\, for cooler stars.  The T Tauri
types of the components within these systems are discussed in \S 3.2.2.

The primary star, A, of each binary pair is defined to be the more massive
component, if mass estimates based on spatially resolved optical
measurements are available (\S 3.2).  For systems without resolved
optical measurements, the primary is assumed to be the component that
is brighter at $2.2\, \mu$m.  This assignment of primary and secondary
is consistent with previous studies of these systems, except in the
case of LkHa 332/G1.  As will be shown in \S 3.2, the fainter component of
this system (historically denoted as 'B') appears to be the more massive
star.  To avoid confusion in comparisons with previous measurements, the
'A' component of this system will always refer to the brighter component
while the 'B' component is the true primary.

\subsection{Near-Infrared Imaging}

\subsubsection{Observations}

Observations at $K$ ($2.2\, \mu$m) and $L$ ($3.6\, \mu$m) of Taurus-Auriga
binaries were obtained on 1996 Dec 5-6 and 1997 Dec 4-9 using the NASA
Infrared Telescope Facility in Hawaii with NSFCAM \citep{rayner93,
shure94}, which contains a $256 \times 256$ array.  All images were
obtained using the finest plate scale available.  The plate scale
(0\farcs0532 $\pm$ 0\farcs0010 / pixel) and orientation with respect to the
cardinal directions ($0.0^\circ \pm 1.0^\circ$) were established through
observations of known
binaries.  Systems with separations less than 1\farcs4 were observed in a
speckle imaging mode.  These observations were conducted using only the
central $128 \times 128$ pixels of the detector (for faster readout times),
resulting in a field-of-view of 6\farcs8 $\times$ 6\farcs8.  Typically,
1200 snapshots (0.1 second exposures) in sets of 400 were obtained for each
source, interleaved with similar observations of a point source.
Photometric standards were also observed regularly in the same manner
for photometric calibration, weather permitting.  Direct images of
systems with separations greater than 1\farcs4 were obtained on 1997
Dec 6 during photometric conditions using the full $256 \times 256$
array of NSFCAM, resulting in a field-of-view of 13\farcs6 $\times$
13\farcs6.  For each pair, dithered images with a total on-source
integration time of 25-200 seconds were obtained, depending upon
source brightness.  Photometric standards and the single star SAO
76599 \citep{cs97} were also observed for photometric calibration and
point spread function (PSF) fitting.

\subsubsection{Data Analysis \& Results}

All the NIR images are first sky subtracted, flat fielded and bad pixel
corrected.  The speckle images are subsequently analyzed with both
classical speckle analysis \citep{labeyrie70} and a shift-and-add technique
\citep[e.g.,,][]{bc80, christou91, ghez98}.  The speckle analysis yields the
diffraction-limited Fourier amplitudes of each object.  The binary star
flux ratio, separation, and orientation ($\pm 180^\circ$) are determined
from a $\chi^2$ minimization of a two-dimensional fit to the Fourier
amplitudes (see Ghez et al. 1995 for details of the fitting procedure and
the error analysis).  The remaining $180^\circ$ ambiguity in the position
angle is eliminated on the basis of the shift-and-add images.  The rapid
sampling of the speckle imaging mode provides a useful check of the
photometric conditions.  When the count rate is constant over the
$\sim$ 1200 frames, the observations are calibrated relative to the
photometric standards.   These results are reported in Table \ref{tab_nir}.

The direct images for each of the wider pairs are registered
and averaged to produce a high signal-to-noise ratio image.  Photometry for
the individual components of the multiple star systems is then obtained
from these average images in two steps.  First, the photometry of each
system is determined by comparing the flux densities measured in large
apertures to the flux densities of
the photometric standards.  Next, the relative flux densities 
and positions of each pair are determined by PSF fitting using the
DAOPHOT package within IRAF\footnote{IRAF is distributed by the
National Optical Astronomy Observatories, which are operated by the
Association of Universities for Research in Astronomy, Inc., under
cooperative agreement with the National Science Foundation.}.
Uncertainties in the system's magnitudes and flux ratios are calculated
from the rms variations of the results from carrying out the same
procedure on the original unregistered images.  The uncertainties
associated with the relative positions are estimated by combining the
rms variations of the results from the unregistered images with the
uncertainty in the plate scale and array orientation.  System
magnitudes and flux ratios for the pairs imaged directly are combined
and the resulting component magnitudes and relative positions for each
pair are listed in Table \ref{tab_nir}.  

The new speckle and direct imaging results provide the first spatially
resolved $L$-band measurements for 32 of the 41 pairs resolved at both $K$
and $L$.  Since the new $K$ and $L$ speckle observations were not
obtained simultaneously and often not photometrically, the results are
combined with previous measurements to obtain component magnitudes.
This is done by using the mean flux density \citep[typically taken
from the compilation of][]{kh95} and median flux ratio for each system.
The resulting
magnitudes for each component are given in Table \ref{tab_nir},
following any new observations reported here.  Uncertainties in the
component magnitudes are based on the standard deviations of multiple
measurements, when available, and therefore should reflect the variability
of the system as well as the measurement uncertainties; the references used
are listed.  Additionally, $K$ and $L$ measurements from the literature of
Haro 6-10 and HV Tau \citep{lh89, menard93, wl98}, which were not observed
here, are combined and listed in Table \ref{tab_nir}.  Although many of the
combined observations are non-simultaneous, this is not likely to
compromise our interpretation since the variations of most young stars at
NIR wavelengths tend to be relatively small ($\sigma_{K}$ $<$ 0.2
magnitudes), and variations of their near-infrared colors are even smaller
\citep[$< 0.1$ magnitudes;][]{kh95}.  The $K$ and $L$ component magnitudes
used in the analysis (\S 3 and \S 4) are marked with asterisks in Table
\ref{tab_nir}.

\subsection{Optical and Ultra-Violet Imaging}

\subsubsection{Observations}

Optical and UV measurements of 21 pairs plus 2 newly discovered components
were obtained with the HST (Table \ref{tab_sample}).  These observations
were
initiated in Cycle 4 (program ID 5395) of the HST mission (1994) with a
pilot study of 4 CTTS and 2 WTTS systems \citep{gws97}.  Here we present
the results of a more extensive follow-up study, which was carried out in
Cycle 6 (program ID 6735) of the HST mission (1997-1998) and which focused
specifically on close CTTS binary systems.  The systems observed in Cycle 6
were imaged through the same 5 broad-band filters used in Cycle 4: F336W,
F439W, F555W, F675W, and F814W \citep[filters that are the approximate
equivalents to the Johnson-Cousins $U$, $B$, $V$, $R_c$ and $I_c$
passbands;][]{biretta96}, as well as the narrow-band H$\alpha$ filter
F656N.  In a total of 9 Cycle 6 pointings, 12 close and 1 wide binary
stars were observed.  For each pointing, the primary source was
centered on the Planetary Camera of the WFPC2, which is an $800 \times
800$ CCD with a pixel size of 0\farcs0455 and a field-of-view of
35\arcsec $\times$ 35\arcsec\, \citep{biretta96}.  To improve the
spatial resolution, the sources were imaged at two positions, offset
by 10.5 pixels along the axes of the detector.  To maximize the
signal-to-noise ratio while allowing for possible source variability,
the exposure times were set to achieve half-well counts in the CCD
detector, based on ground-based measurements of the unresolved
pairs.  Specific details regarding these observations can be accessed
from the Hubble Data Archive at the Space Telescope Science Institute
(STScI).

In addition to the pairs observed with the HST, 8 other Taurus-Auriga
binaries have spatially resolved optical measurements and are included in
this study.  Six wide binaries (Haro 6-37 Aa-B, UX Tau A-B, UX Tau A-C, 
HN Tau A-B, V710 Tau A-B, HV Tau A-C) have been resolved with direct
imaging \citep{hss94, kh95, mm94} and 2 close binaries (DD Tau and T
Tau) were resolved with high resolution imaging techniques \citep{btc92,
gorham92, stapelfeldt97}, although T Tau B was not detected.  Together
with measurements from the literature, the new HST observations provide
spatially resolved measurements at optical wavelengths of the complete
sample of CTTS binaries brighter than $V$ = 17 magnitudes\footnote{Haro
6-28 \citep[V = 17.30 mag;][]{kh95} was the only known CTTS binary excluded
because of its magnitude.} and with separations ranging from 0\farcs07 to
0\farcs72, that were known at the time the HST observations were proposed.

\subsubsection{Data Analysis \& Results}

The STScI carried out the initial data
processing through the `calibration pipeline' \citep{holtzmann95a}.  We
further correct the data for the charge transfer efficiency problem
associated with the WFPC2 detector using the relations of \citet{whc99}.
Since the 6 sources observed in Cycle 4 were initially analyzed before
these corrections were available \citep{gws97}, these measurements are
updated with the new calibrations.  Although the majority of charge
transfer efficiency corrections are less than 5 percent, some corrections
for fainter systems observed through the F336W and F439W filters are 10
percent or more and dominate the photometric uncertainty.  Cosmic ray
events are removed by interpolating from neighboring pixels.  Each set of
images is registered and combined by centroiding on the brightest point
sources.

Photometry for the widest components ($> 0\farcs5$) of the multiple star
systems is carried out by summing the counts within 5 pixel (0\farcs23)
radius apertures centered on the stars.  The flux ratios are computed from
the ratios of the components' counts.  System magnitudes in the Vega
magnitude system are determined from the sum of the components' counts,
using the known filter zero points and aperture corrections
\citep{holtzmann95b, keyes97}.  Since the systems observed here typically
have large source counts, the dominant source of uncertainty, in most 
cases, is the uncertainty in the charge transfer efficiency correction,
which we estimate to be one half of the applied correction \citep{whc99}.
The photometric uncertainties are determined by convolving this uncertainty
with the ranges of the flux measurements from the original unregistered
images.

Photometry for the individual components of the closest ($< 0\farcs5$)
multiple star systems is accomplished in two steps.  First, the Vega
magnitudes of each system are determined from the total counts within
35 pixel (1\farcs59) radius apertures and the known filter zero points and
aperture corrections.  Uncertainties are computed in the same manner
discussed above for the wider pairs.  Next, the relative flux densities are
determined using the PSF fitting routines of the DAOPHOT package within
IRAF.  The PSFs used in the fitting process are constructed from either
isolated sources within the images (e.g., UZ Tau A) or from the PSF Archive
available at the
STScI.  These PSFs are typically the average of three or more single stars
that are positioned similarly on the chip as the target binary.  The rms
variations of the results from carrying out the same procedure on the
original images using multiple PSFs provide estimates of the flux ratio
uncertainties.  The flux ratios and system magnitudes for all pairs imaged
with the HST are reported in columns 3 and 4 of Table \ref{tab_wfpc2_bin}.

This photometry procedure works well for all systems expect FS Tau.
Extended emission in this system biases the stellar flux densities
measured in large apertures as described above toward brighter values.
Therefore the component magnitudes for this system are determined directly
by using flux calibrated PSFs in the PSF fitting process.  These values
are combined into a flux ratio and total stellar magnitude and are listed
in Table \ref{tab_wfpc2_bin}.  These values may nevertheless still be
biased because of the extended emission.

Columns 8 and 9 of Table \ref{tab_wfpc2_bin} list the separations and
position angles of the binary stars imaged with the HST.  These are
derived from the photometric centers of the components for pairs wider than
0\farcs5, and from the results of the PSF fitting for closer pairs.  The
uncertainties associated with the relative positions are estimated by
combining the rms variations of the results from the five broad-band
filters with the uncertainty in the plate scale ($\sim 10^{-5}$
arcseconds per pixel) and spacecraft orientation ($\sim$ 0.03$^\circ$)
\citep{holtzmann95a,biretta96}.

Flux ratio lower limits are also given in Table \ref{tab_wfpc2_bin} for the
non-detections:  FW Tau C in the F336W, F439W, and F555W filters and FV
Tau/c B in the F336W and F439W filters.  Since these two stars are
detected through at least two filters, these positions are used to
identify their locations in the other images.  A scaled PSF is added at
these positions to determine the minimum detectable flux densities using
the DAOPHOT package within IRAF.  No companion is detected to the
`suspected binary' FY Tau \citep{richichi94}.  This missing companion is
discussed below.

With system magnitudes and flux ratios, the magnitudes of each binary
star component are determined.  In order to better compare the HST
broad-band measurements with previous ground-based observations and
standard photospheric colors, the HST broad-band measurements of each
resolved star are transformed to the more standard Johnson-Cousins
filter system.  These filter transformations are derived in Appendix
A.  The transformation values are typically small, with only weak
dependencies on stellar temperature (spectral type) for stars
hotter than about M5.  The spectral types inferred in \S 3.1
are used to assign specific transformation values.
Magnitudes in the Johnson-Cousins filter system for the T Tauri stars
observed here are listed in columns 6 and 7 of Table \ref{tab_wfpc2_bin}
and are used in the analysis of the stellar and circumstellar properties
(\S 3 and \S 4).  The sums of the components' flux densities for all pairs
agree well with the ranges measured in previous lower resolution
ground-based measurements \citep[e.g.,][]{herbst94}.

The measured F656N flux densities ($F_{obs}$) are converted to
EW[H$\alpha$] measurements, a common measure of H$\alpha$ emission.  In
order to estimate and remove the photospheric contribution to the
narrow-band flux, F656N - F675W colors are derived from dwarf standard
stars that have no appreciable H$\alpha$ emission
(Appendix A).  The observed F675W flux densities, attributable primarily to
photospheric emission, are then converted to photospheric F656N flux
densities ($F_{phot}$) using these colors.  Since the F656N filter has an
effective rectangular width of $28.3$ \AA\, and is centered on H$\alpha$ 
\citep{biretta96}, the EW[H$\alpha$] values are approximated as
EW[H$\alpha$] = $-28.3$ \AA\, $\times (F_{obs} - F_{phot}) / F_{phot}$ and
are listed in Tables \ref{tab_wfpc2_bin} and \ref{tab_wfpc2_fs}.
Uncertainties are estimated from the convolution of the flux density
uncertainties, the transformation uncertainty (0.02 magnitudes), and the
filter width uncertainty (1 \AA).  Combining the flux
densities for the components results in photometric EW[H$\alpha$] estimates
for the systems that are in reasonable agreement with previous
spectroscopic measurements of the unresolved pairs \citep[e.g.,][]{hb88}.
Specific differences are most likely because of variable
H$\alpha$ emission, a common characteristic of classical T Tauri stars.
However, for the strongest H$\alpha$ emission sources (EW[H$\alpha] > 100$
\AA), the photometric estimates provide only lower limits since the
emission-line will begin to contribute significantly to the F675W
measurements.  The photometric method for estimating EW[H$\alpha$] is
therefore, at the very least, sufficiently robust for distinguishing
between CTTS and WTTS types.

\subsubsubsection{Missing Companions}

The suspected companion to FY Tau \citep[separation $>$ 0\farcs
15;][]{richichi94} is not detected in the HST images.  Likewise, as
discussed in \citet{gws97}, the proposed companion of RW Aur B
\citep{gnm93} is not detected in the HST images.  These suspected
companions have only been reported once based on low signal-to-noise ratio
NIR measurements.  If the NIR measurements in
question are used to estimate the photospheric flux at $I_c$ (assuming no
$K$ excess emission, the same age and extinction as the primary, and dwarf
colors, see \S 3.1), the predicted fluxes are 4.9 magnitudes (RW Aur C)
and 4.2 magnitudes (FY Tau B) brighter than the $I_c$ detection limits
($\sim 17.8$ magnitudes in both cases).  Under these assumptions, these
companions should be easily seen in the HST images.  It is possible that
these missing companions are substantially more extincted than their
primary, but the considerable extinction differences required to reconcile
the colors ($A_V$[sec] $> A_V$[prim] $+$ 8.0 magnitudes) are
inconsistent with the similar extinctions of binary star components (\S
3.1).  We consider it more likely that the weak lunar occultation detection
of FY Tau B was caused by scintillation, as considered possible by
\citet{richichi94}.  The similarly weak detection of RW Aur C may also have
been a false detection.  We, hereafter, consider FY Tau to be a single star
and RW Aur B to be a single star in a wide binary.

\subsubsubsection{New Companions}

New companions are discovered in the V410 Tau system \citep[see][]{gws97}
and the FW Tau system in the HST images.  The new companion to FW Tau,
at a separation of 2\farcs 3, is particularly intriguing as it is
exceptionally bright in the narrow-band H$\alpha$ filter, but is very faint
in the F675W and F814W filters and is not detected at shorter wavelengths.
The strong Balmer series emission (EW[H$\alpha$] $> -454 \pm 224$; Table
\ref{tab_wfpc2_bin}) is
typical of non-stellar Herbig-Haro objects, which are knots of shocked gas
that typically lie along the path of stellar jet \citep{schwartz83}.
However, the emission comes from an isolated, unresolved point source and
the nearby WTTS binary FW Tau A \& B is an unlikely source for a jet.
Since FW Tau C is detected through the F814W filter, which does not
include many strong emission-lines commonly found in Herbig-Haro objects
\citep{schwartz83}, a stellar interpretation is preferred.  

The single
star FZ Tau, at a separation of 17\farcs2 from the suspected binary FY
Tau (see above), was serendipitously observed but is not considered a
companion.  For completeness, the measurements of the single stars FY Tau
and FZ Tau are provided in Appendix B.

\subsubsubsection{Extended Emission}

With the exception of FS Tau, none of the systems show evidence of extended
nebulosity.  Since the exposure times for all sources were scaled similarly
to avoid saturation (\S 2.3.1), the sensitivity to extended emission is
limited, but is nevertheless roughly uniform for the entire sample.
Limiting surface
brightnesses are estimated from the standard deviation of the counts per
resolution element (0\farcs09 $\times$ 0\farcs09 or 2 $\times$ 2 pixels) at
a separation of 0\farcs5 from the primary star's position in the residual,
PSF subtracted images (\S 2.3).  Integrated over 0.008 arcsec$^2$, the
surface brightness is typically 7 magnitudes at F555W and 8 magnitudes at
F814W below the magnitude of the primary.  All but one (19/20) of the T
Tauri star systems imaged with the HST in this program appear to be
unresolved point sources at this resolution and dynamic range.

In contrast to the other systems, the FS Tau system reveals extended
emission at all wavelengths observed with the HST.  This is also seen in
the long-exposure images obtained by \citet{krist98} with HST through the
F555W and F814W filters.  With our non-saturated images over a broader
range of wavelengths,
we analyze FS Tau's extended emission by differencing the results of
wide aperture photometry and the stellar flux densities derived above.  In
order to compare the results directly with those of \citet{krist98}, a
radius of 4\farcs0 is used for the wide aperture photometry.  A spectral
type of M1, which was measured for the entire system by \citet{ck79}, is
adopted for the filter transformations.  The resulting values are listed in
Table \ref{tab_wfpc2_fs} and, in the case of the wide aperture magnitudes,
are fainter by 0.15 magnitudes at both $V$ and $I_c$ than the earlier
measurements.  Nonetheless, like the \citet{krist98} results, the nebulous
emission appears bluer than the stellar emission ($V-I_c$[Nebula] =
2.52 versus $V-I_c$[Binary] = 3.19) as is expected for scattering by small dust
grains.  There is no evidence of H$\alpha$ emission from the
nebulosity as might be expected from the interaction of a stellar jet with
remnant cloud material \citep{em98}.  The spatial distribution of the
extended emission in the narrow-band H$\alpha$ image is the same as in the
broad-band images, as expected for dust scattering.  Additionally, the
EW[H$\alpha$] of the nebulosity ($-40.8 \pm 8.0$ \AA) is actually
diminished relative to the stellar pair ($-72.1 \pm 3.5$ \AA), suggesting a
higher optical depth at H$\alpha$ due to H I gas in the nebulosity.

\subsection{Optical Spectroscopy}

\subsubsection{Observations}

The Faint Object Spectrograph (FOS) aboard the HST (program ID 6014)
was used to obtain spatially separated spectra of the close binaries
XZ Tau (1996 Feb 2), GH Tau (1996 Oct 6) and V955 Tau (1996 Dec 4).
GG Tau A was also observed (1995 Nov 8) as part of this program, and
its spectra were presented in \citet{white99}.  These observations
were conducted using the 0\farcs09 $\times$ 0\farcs09 aperture with
the G570H (4600 - 6800 \AA) grating, which yielded a spectral
resolution of R $\approx$ 1400 \citep{keyes95}.  Other observational
details can be accessed from the Hubble Data Archive at STScI.
The observations were carried out by first centering the 0\farcs09
aperture on the brighter component, assumed to be the primary, and
obtaining its spectrum and then offsetting to the known location of the
secondary.  This strategy worked well for GG Tau A and spectra of both
components were obtained.  For the XZ Tau system, the
companion star was unexpectedly the brightest optical component at the
time of the observations and consequently its spectrum was obtained first.  
The second spectrum obtained at the nominal offset resulted in a spectrum
of the sky on the opposite side of the primary star.  To avoid this
incorrect source acquisition due to source variability again, the FOS
acquisition strategy for GH Tau and V955 Tau was adjusted to obtain an
offset spectrum at both plus and minus the separation of the companion.
This strategy resulted in spectra of both components of the GH Tau system
as well as a spectrum of the sky.  While the measurement of V955 Tau A's
spectrum was successful, the attempt on V955 Tau B failed due to an
unusual telescope error that offset the telescope to the wrong position.
Altogether, these observations resulted in spatially resolved spectra for 6
components.

\subsubsection{Data Analysis \& Results}

All HST spectra were initially calibrated by the FOS calibration
pipeline \citep{keyes97}.  However, flat fielding information for small
apertures was not initially available, thus the spectra are flat fielded
using the appropriate sensitivity functions provided by T. Keyes \&
E. Smith (private communication), which are now available in the
calibration pipeline.  The resulting spectra of XZ Tau B, GH Tau A \& B
and V955 Tau A are shown in Figure \ref{fig_fos} (see White et al. 1999 
for GG Tau A results).  The spectra of all components show strong
H$\alpha$ emission typical of classical T Tauri stars (EW[H$\alpha$] =
$-76 \pm 4$ \AA\, for XZ Tau B, $-12 \pm 1$ \AA\, for GH Tau A, $-15 \pm 1$
\AA\, for GH Tau B, $-18 \pm 2$ \AA\, for V955 Tau A).  Because of the
limited resolution, no Li I at 6708\, \AA\, is detected in any of the FOS
spectra; approximate equivalent width upper limits to the detection of any
features are 1 \AA.  Of particular interest is the spectrum of XZ Tau B,
which exhibits prominent Balmer series emission (H$\alpha$ and H$\beta$)
and an array of strong emission-lines, including Fe I, Fe II, and Mg I
features.  The spectrum and emission-line intensities are remarkably
similar to those seen in the spectrum of DG Tau, a very high accretion
single T Tauri star with prominent jet \citep{hg97}, and suggests that XZ
Tau B powers the jet in this system \citep{krist99}.

The primary goal of the spectroscopic analysis presented here is accurate
spectral type classification of each component.  Spectral types are
established by comparison with spectral standards from \citet{montes97}
over the temperature sensitive region 5700 - 6800 \AA.  This longer
wavelength portion of each spectrum is used for spectral classification
since it suffers the least from continuum excess emission common in
classical T Tauri stars \citep{bb90, hartigan91}.  The
spectral types inferred from these comparisons are listed in Table
\ref{tab_bin_stel}.  The best fit dwarf spectra are also shown in Figure
\ref{fig_fos}.

While the majority of late-type T Tauri stars can be assigned spectral
types with moderate resolution spectra, some T Tauri stars display a
substantial continuum excess and strong emission-lines which can
compromise the accuracy of this classification (e.g., XZ Tau B).  The
resolution of the FOS
spectra is insufficient to determine the level of continuum excess
emission, or veiling, which may be present in the spectra.  This typically
requires spectroscopy of sufficient spectral resolution (R $> 10,000$) to
resolve individual atomic features.  The strength of Balmer series
emission, however, is correlated with continuum veiling \citep[see \S
3.2.1;][]{bb90}.  Since GH Tau A, GH Tau B, and V955 Tau A all show
relatively weak H$\alpha$ emission ($<$ 20 \AA), at levels only modestly
above the
WTTS limit (\S 2.1), it is likely that they experience little or no
optical veiling.  In contrast, XZ Tau B clearly exhibits a continuum
excess, although based on the available spectroscopic information there is
no definitive way to accurately quantify this.  Only the reddest portion of
the spectrum shows any photospheric features.  Both CaH absorption (6382
\AA\, \& 6389 \AA) and the step-like features of TiO ($>$6600 \AA) are
visible and indicate an underlying photosphere.  Nevertheless, the
depths of these temperature sensitive features are likely diminished
because of continuum excess emission.  While this region of the spectrum
appears somewhat similar to the M1 standard shown in Figure \ref{fig_fos},
this spectral type is more likely an upper limit with respect to
temperature.

\section{Binary Star Property Results}

\subsection{Stellar Properties}

\subsubsection{Temperature, Extinction, and Luminosity Estimates}

The standard method for deriving stellar temperatures and luminosities
relies on both a spectroscopically derived spectral type and broad-band
optical photometry.  The latter is measured in this study for the
component stars of the 25 systems described in \S2.3.  Only 8 of these
systems, however, have spatially separated spectra and hence spectral types
known for both components.  For the remaining systems, the optically
inferred spectral type of each unresolved system is assigned to the
component that is the brightest at optical wavelengths.  This is the primary
in all cases except for LkHa 332/G1.  Uncertainties of one spectral
subclass are assumed for all spectral types unless otherwise noted (Table
\ref{tab_bin_stel}).  These spectral types are converted to temperatures
using a temperature scale that is consistent with dwarf temperatures for M0
and hotter stars \citep{bb88} and is moderately hotter than dwarf
temperatures for cooler stars \citep{luhman99}.  In order
to derive stellar luminosities, it is first necessary to estimate
extinction values.  These values are determined from the extinction law of
\citet{rl85} and a comparison of the $V-I_c$ colors observed with those
corresponding to the from the spectral types, using the dwarf color
relations of \citet{bb88} and \citet{bessell91} for K7 and hotter stars and
that of \citet{km94} for M0 and cooler stars.  Uncertainties in the
extinctions are estimated by changing the spectral types by their
uncertainties and rederiving the extinctions.  Finally, the luminosities
are derived from the dereddened $I_c$ magnitudes, using a distance of 140
pc \citep{kenyon94, preibisch97}.  The $I_c$ bandpass is chosen for
calculating the stellar luminosities since it suffers the least
contamination from UV excess.  Bolometric corrections from
\citet{bessell91} are used for stars hotter than spectral type M3 and
values from \citet{monet92} are used for spectral types of M3 and cooler.
While bolometric corrections for early M stars are reasonably well
established, their remains considerable uncertainty in the bolometric
corrections for late-M stars \citep{monet92, kirkpatrick93}.  The
uncertainties in the adopted bolometric corrections are assumed to be 0.05
magnitudes for spectral types hotter than M5, 0.1 magnitudes for spectral
types between M5 and M7, and 0.2 magnitudes for spectral types of M7 and
cooler.  These uncertainties are combined with the uncertainties in the
photometry and the extinction estimates in order to determine the
uncertainties in the stellar luminosities.  The inferred properties are
listed in Table \ref{tab_bin_stel}.

For the companion stars without spectroscopic measurements of their
spectral types, we assume that the extinction to the secondary is the same
as to the primary.  This assumption is checked by comparing the
independently derived extinctions for the components of 14 fully resolved
(photometrically and spectroscopically) T Tauri star systems with
separations $<$ 1000 AU, and with no known unresolved companions (Fig
\ref{fig_bin_av}).  The difference in extinction
between the primary and secondary stars for these systems is consistent
with no difference: mean [$A_V$(prim) - $A_V$(sec)] =
-0.15 magnitudes, with a standard deviation of $0.80$
magnitudes.  The standard deviation of these differences is reduced
considerably if the most discrepant case, GG Tau Aa and Ab, is ignored:
mean [$A_V$(prim) - $A_V$(sec)] = 0.01
magnitudes, with a standard deviation of $0.54$ magnitudes.  Proceeding
with the same extinction assumption, these extinction estimates are used to
deredden the observed $V-I_c$ colors of the secondaries.  The spectral types
and corresponding temperatures of these stars are then estimated from
their reddening corrected $V-I_c$ colors\footnote{Although the
transformations used to determine the Johnson-Cousins photometry depends on
the spectral types (\S 2.3.2), this dependence is very weak and the
solutions can be determined in one iteration.}.  For GG Tau Bb, HN Tau B,
and UX Tau C, $R_c-I_c$ colors are used since these stars have no $V$-band
measurements.  The uncertainties in the spectral types of the secondaries
are estimated by changing the spectral types of the primaries by their
uncertainties and then rederiving the extinctions and the colors of the
secondaries; a minimum uncertainty of 1 spectral subclass is assumed.  The
luminosities of the secondaries are derived from the dereddened $I_c$-band
magnitudes as described above.  The stellar properties for these companion
stars are also listed in Table \ref{tab_bin_stel}.

A few of the components measured photometrically have additional optically
unresolved companions that need to be addressed individually.  For the
single-lined spectroscopic binaries UZ Tau E \citep{mathieu00} and RW Aur A
(Gahm et al. 1999, but see also Petrov et al. 2001), all the observed flux
is attributed to the primary.  For
the double-lined spectroscopic binary V773 Tau A, the primary's flux is
estimated from the flux ratio at R and the observed spectral types
\citep{welty95} as described in \citet{gws97}.  Four binary components -
V807 Tau B, HV Tau A, Haro 6-37 A, and UX Tau B - each have an additional
companion that has been resolved in NIR lunar occultation or speckle
imaging measurements \citep{duchene99a, gnm93, simon92}, but have no
spatially resolved optical measurements.  In order to correct for the
emission from this companion, the magnitude difference at $J$ (1.25
$\mu$m), if available (Haro 6-37 A and UX Tau B), or $K$ is used to
determine the approximate temperature and spectral type of the companion.  
First, the unresolved pair's spectral type and corresponding temperature
are assigned to the brighter star.  Then the companion's temperature is
estimated from the absolute magnitude - temperature relations of the
\citet{baraffe98} evolutionary models (\S 3.1.2), at an age of 3 Myrs.
Finally, the companion's contribution to each bandpass is then estimated and
removed using the spectral type - color relations from the references
listed above, normalized to the observed $J$ (or $K$) magnitude.  Only the
flux corrected brighter star of each pair is included in the subsequent
analysis.

For comparison with the components of multiple star systems, single stars
in Taurus-Auriga are identified in Appendix B.  The stellar properties of
these stars are derived following the same procedure used for stars with
both spectra and photometry, and these properties are reported in Table
\ref{tab_sing}.

While the adopted methodology for determining luminosities, temperatures,
and extinctions works well for the majority of T Tauri stars, this method
is inadequate if the optical light is not predominately from the stellar
photosphere.  Young stars with exceptionally high levels of accretion are
examples of this.  In these cases, the optical light is dominated by
continuum excess emission from the accretion shock
\citep[e.g.,][]{gullbring00}.  This is illustrated in the spectra of
several component stars studied here such as XZ Tau B (Fig \ref{fig_fos}),
HV Tau C \citep{mm94}, and HN Tau A \citep{mmd98}; the spectra show
no or barely discernible photospheric features.  An additional property
that these
high accretion stars have in common is that they are exceptionally red at
near- and mid-infrared wavelengths.  Since we have $K-L$ colors for the
majority of binary star components studied here (Table \ref{tab_sample}),
we use the well known correlation between $K-L$ color and optical excess
emission (\S 3.2.1) to identify stars with abnormally high optical excess
levels.  A linear fit to this relation implies that stars with $K-L$ colors
$>$ 1.4 magnitudes have optical excess emission levels that are $>$10 times
that from the stellar photosphere (\S 3.2.1).  This high
level of excess emission is greater than the levels spectroscopically
determined for all the CTTSs in \citet{heg95}, and thus we label stars with
$K-L$ colors above this value as \textit{high accretion stars}.  This
classification is further supported by optical spectroscopy.  All
stars with $K-L$ colors greater than 1.4 magnitudes display nearly
featureless optical spectra, when available, while those with bluer $K-L$
colors typically show well defined photospheric features and consequently
have better established spectral types.  The list of high accretion stars
includes the binary star components CZ Tau B, FS Tau A, FS Tau B, Haro 6-10
B, HV Tau C\footnote{Although HV Tau C has not been detected at $L$, it is
classified as a high accretion star because of its featureless optical
spectrum and its exceptionally red $K-N$ color \citep{mm94, wl98}.}, T Tau
B, XZ Tau B, FV Tau/c B, and HN Tau A, and the single stars DG Tau, DR Tau,
and HL Tau.  Since the inferred 
luminosities and temperatures for these high accretion stars may be in
error, the derived values are listed in Tables \ref{tab_bin_stel} and
\ref{tab_sing} in parentheses.

\subsubsection{Absolute Age and Mass Estimates}

Masses and ages of the binary star components are estimated by comparing
the stellar luminosities and temperatures with the predictions of
pre-main-sequence (PMS)
evolutionary models.  Although considerable uncertainties remain in the
assumptions used to calculate these models, recent studies by
\citet{white99} and \citet{sdg00} have shown that the evolutionary models
of \citet{baraffe98} computed with a mixing length equal to nearly twice
the pressure scale height are consistent with the available observational
constraints.  The \citet{baraffe98} models, however, are only computed for
masses between 0.025 M$_\odot$ and 1.20 M$_\odot$.  In order to extract
masses and ages for the entire range of masses in this study, the low mass
evolutionary models of {\citet{baraffe98} are combined with the higher mass
models of \citet{ps99}, as was done by \citet{white01}.  These models agree
reasonably well at the adopted transition mass of 1.00 M$_\odot$.

In Figure \ref{fig_bpsbin}, the components of binary systems with spatially
resolved optical measurements are plotted on an H-R diagram along with the
adopted models for
PMS evolution.  Single T Tauri stars (Appendix B) are also shown, and high
accretion stars (both binary components and single stars) are distinguished
and labeled.  The inferred masses and ages for the binary star sample are
listed in Table \ref{tab_bin_stel}.  Uncertainties are set by the ranges of
masses and ages inferred by changing the temperature and luminosity
estimates by their uncertainties (Table \ref{tab_bin_stel}).
Stars with a luminosity above the youngest isochrone are assigned an age
upper limit of 1 Myrs.  The ages derived for these binary star components,
excluding the high accretion stars, range from $< 1$ Myrs to 19 Myrs, with
a mean log[age] of $6.43 \pm 0.05$ dex ($\sim 2.7$ Myrs; includes 45 stars,
$\sigma = 0.35$ dex).  The masses range from 0.042 M$_\odot$ to 2.11
M$_\odot$,
with only one component, GG Tau Bb, below the hydrogen burning minimum mass
\citep[i.e. a brown dwarf;][]{white99}.  These values are consistent with
the overall population of single T Tauri stars in Taurus-Auriga (Appendix
B).  The combined sample of singles and binary components have a mean
log[age] of $6.44 \pm 0.04$ dex (includes 94 stars, $\sigma = 0.37$ dex).
The non-physical luminosity and temperature estimates derived for the high
accretion stars are illustrated by their
systematically under-luminous locations on the H-R diagram relative to the
majority of T Tauri stars.  Since the optical measurements for these stars
are insufficient to determine their stellar properties, it is not possible
to confidently determine which component is the true primary (i.e., more
massive).  Consequently, these objects are excluded from the following
discussion of relative masses and ages, but are revisited in \S 4.3.

\subsubsection{Relative Ages and Masses of Binary Star Components}

The component ages listed in Table \ref{tab_bin_stel} are used to
test the coevality of binary star components.  GG Tau B is omitted from
this analysis since it was used to define the temperature scale with the
assumption of coevality imposed on it \citep{luhman99, white99}.  As shown
in Figure \ref{fig_agesep}, the components of binary stars are, on average,
consistent with being coeval.  If the four components with age upper limits
are assigned an age of 1 Myrs (1 primary star and 3 secondary stars), the
differences in the components' log[age] have a mean of $0.07 \pm 0.07$ dex
and a standard deviation of $0.31 \pm 0.05$ dex.  This standard deviation
is comparable to the average uncertainty in the ages ($\sim$
0.30 dex).  We therefore conclude that the observed spread in relative ages
is dominated by measurement uncertainties, and estimate a 3$\sigma$ upper
limit to the intrinsic log[age] differences to be 0.15 dex. For a typical
T Tauri star age (3 Myrs; \S 3.1.2), this implies that binary stars are
coeval to $\sim$ 1 Myrs.  Divided into two sets, the close and wide pairs,
this sample shows that close pairs have a slightly smaller spread in age
compared with wider pairs ($\sigma = 0.27 \pm 0.05$ dex versus $\sigma =
0.39 \pm 0.11$ dex), but the difference is not statistically significant.

An additional test of coevality that is less dependent on accurately
converting measurement uncertainties to age uncertainties is a comparison
of the relative ages of the binary star components and randomly paired
single T Tauri stars.  An ensemble of random pairings of 49 single T Tauri
stars in Taurus-Auriga (Appendix B) leads to a median standard deviation in
log[age] of 0.54 dex, which is considerably larger than that of the binary
sample ($0.31 \pm 0.05$ dex).  An example set of randomly paired single
stars is shown in Figure \ref{fig_agesep} for comparison with the binary
star sample.  Although the log[age] differences of the binary star sample
appear to be significantly less that the log[age] differences of the
randomly paired single star sample, effects other than a larger spread of
intrinsic ages could, in principle, produce apparently larger age
differences.  For example, larger measurement uncertainties for the single
stars could have this effect.  However, the stellar properties of the single
stars are actually more accurately determined since they all have spectral
types from spectra and multi-epoch photometry.  Uncertainties in the
evolutionary models and temperature scale could also produce a larger age
spread, and this effect would be the most significant for pairs with the
largest differences in temperature.  If binary stars and randomly paired
single stars with similar temperature differences are compared, however,
the binary stars are more coeval than the single stars over all ranges of
temperature differences.  Another possibility is that our method of
equating the $A_V$'s for some of the binary star components biases the
binary star pairs toward similar ages.  To check this, we analyze
the sub-set of the binaries for which both components have been measured
spectroscopically and photometrically.  This set has a standard deviation
of $0.34 \pm 0.08$ dex, similar to the whole binary star sample, suggesting
that the method has not affected our analysis.  We therefore conclude that
the components of T Tauri binary stars are significantly more coeval than
randomly paired single T Tauri stars within the same star forming region.

The component mass estimates are used to study the distribution of binary
star mass ratios (m$_\textrm{s}$/m$_\textrm{p}$).  Mass ratios for 21 pairs
with optically resolved measurements (Table \ref{tab_bin_stel}) are plotted
versus binary separations in Figure \ref{fig_hst6_mass} (\textit{top
left}).  As shown in Figure \ref{fig_hst6_mass} (\textit{top right}), the
mass ratios are correlated with the $K$ magnitude differences
between the primaries and the secondaries.  Thus we use this empirical
relation (m$_\textrm{s}$/m$_\textrm{p} = 1 - 0.328\Delta K$)
to estimate mass ratios for the 20 additional binaries with spatially
resolved $K$-band measurements (\S 2.1).  Pairs with a high accretion star
are excluded from both samples.  Figure \ref{fig_hst6_mass} (\textit{bottom
left}) shows that the mass ratio distributions determined from the optical
measurements and from $K$ magnitude differences are similar and are both
peaked near unity.  Combining both samples, Figure \ref{fig_hst6_mass}
(\textit{bottom right}) shows that the majority of high mass ratio pairs
are in close systems.  Here the definition of 'close' is extended
slightly to 110 AU to include FQ Tau; this separation shows the clearest
break in the distribution of mass ratios versus separation (Fig 
\ref{fig_hst6_mass}, \textit{top left}).  A K-S test shows that these
distributions are different at the 95\% level, a 2$\sigma$ difference.
One limiting observational bias is that faint, low mass ratio binaries are
more difficult to detect at close separations than at wide separations.
However, an identical significance (95\%) is
determined if the K-S test is conducted on close and wide pairs excluding
binaries with $K$ magnitude differences of 3.0 or more (1 close, 1 wide).
A similar mass ratio dependence on separation, as traced by $K$ flux ratio,
was reported in the Taurus-Auriga binary study of \citet{kl98}, which
consisted of 85 pairs, 34 of which are X-ray identified systems
\citep{wichmann96} and thus are not included in the sample studied here.
Thus, although multiplicity surveys that are more sensitive to close, low
mass companions are needed to minimize observational biases and confirm
this tentative suggestion, it appears that high mass ratio pairs are
slightly more common among close ($\sim$ 10-110 AU) binaries than among
wide ($\sim$ 110-1000 AU) binaries in Taurus-Auriga.

\subsection{Circumstellar Properties}

Standard accretion disk diagnostics are used to study the circumstellar
disks associated with the components of the binary T Tauri stars studied
here.  Specifically, UV excesses, $K-L$ colors, and the EW[H$\alpha$]s are
used to identify circumstellar accretion disks, and the strengths of these
diagnostics are used to characterize mass accretion rates.  Since 
these accretion disk diagnostics have been discussed extensively in the
literature \citep[e.g.,][]{hartmann98}, in \S 3.2.1 we very briefly
summarize their effectiveness through correlations with
independently determined, optical continuum excesses extracted from high
resolution spectra, a direct tracer of the mass accretion rate
\citep{heg95}.  These relations are explored using the single T Tauri star
sample (Appendix B), which is free of biases introduced by unresolved
companions.  In \S 3.2.2, the T Tauri types (CTTS versus WTTS) of the
binary star components are assigned and the effectiveness and consistency
of the diagnostics used for type classification are discussed.  The
circumstellar properties of the binary star systems are presented in \S
3.2.3.

\subsubsection{Signatures of Circumstellar Disks}

\subsubsubsection{Ultra-Violet Excesses and Mass Accretion Rates}

The accretion of circumstellar material produces substantial
excess UV luminosity due to the dissipation of the accreting material's
kinetic energy as it impacts the stellar surface.  Although this excess
luminosity is most accurately measured from high resolution spectroscopy
\citep[e.g.,][]{bb90}, it can also be confidently inferred from broad-band
UV measurements.  Figure \ref{fig_duklha} (\textit{top left}) shows the
correlation between the excess emission measured in the $U$-band
filter\footnote{The $U$-band excess is quantified as $\Delta U$, which is
2.5 times the logarithm of the ratio of the total $U$-band flux and the
photospheric $U$-band flux, where both quantities are reddening corrected
($\Delta U$ = 2.5$\times$log\,[$F_{tot}/F_{phot}$]).} and the optical
continuum excess emission, or optical
veiling, measured in high resolution spectra \citep{heg95}.  Stars with the
largest $U$-band excesses have the largest levels of optical veiling.  In a
similar but more useful comparison, \citet{gullbring98} quantified the
relation between $U$-band excess measurements and the total accretion
luminosity determined from optical veiling.  Following their empirically
determined relation, we estimate accretion luminosities for single CTTSs
(Appendix B) from $U$-band excesses.  These values are then converted to mass
accretion rates following the model of \citet{gullbring98}, in combination
with stellar radii ($L_\star = 4\pi R_\star^2\sigma T_\star^4$)
and mass estimates.  The accretion rates, listed in Table \ref{tab_sing},
range from $3 \times 10^{-10}$ M$_\odot$ yr$^{-1}$ to $7 \times 10^{-8}$
M$_\odot$ yr$^{-1}$ (mean log[$\dot{\textrm M}$] = -8.33, $\sigma = 0.67$),
and are consistent with the range of accretion rates
inferred from spectroscopic data \citep{gullbring98}.  The limitation of
this method is that the moderate $U$-band excesses common in WTTSs and
attributable to their active chromospheres give the impression of low level
accretion.  The observed $U$-band excesses for the single WTTSs are used to
determine upper limits to their mass accretion rates (Table
\ref{tab_sing}), adopting the same methodology described above.  A mass
accretion rate upper limit of $10^{-10}$ M$_\odot$ yr$^{-1}$ is assumed for
LkCa 19 and LkCa 5, which both have slightly negative $U$-band excesses.
Chromospheric excesses can give the appearance of accretion rates of up to
a few $\times\, 10^{-9}$ M$_\odot$ yr$^{-1}$.

\subsubsubsection{The Near Infrared Color $K-L$}

The amount of NIR excess emission is known to be correlated with the rate
of circumstellar accretion.  Larger accretion rates must dissipate angular
momentum more quickly in the viscous disk and this leads to a warmer, more
luminous disk \citep{bbb88, kh90}.  The increase in disk emission produces
larger NIR excesses and redder NIR colors.  The $K-L$ color is a
particularly useful NIR accretion diagnostic since the underlying
photospheric color exhibits little spectral type dependence ($K-L$[K0] =
0.06 versus $K-L$[M5] = 0.29) and it is not significantly affected by the
moderate extinction ($A_V \sim 1-2$ magnitudes) associated with most T
Tauri stars.  In Figure \ref{fig_duklha} (\textit{bottom left)}, the
observed $K-L$ colors for single T Tauri stars are compared to their
optical excesses.  All sources with $K-L$ colors larger than $\sim$ 0.3
magnitudes have detectable optical excesses, while those with normal
photospheric $K-L$ colors ($\sim$ 0.1 - 0.3) have no detectable optical
excesses.  The $K-L$ color is also correlated with the
level of optical excess emission, and this relation is quantified with a
linear least squares fit\footnote{$K-L$ colors greater than 1.4 imply
continuum excesses that are nearly 10 times that of the stellar continuum,
and this $K-L$ color is therefore chosen to identify high accretion stars
(\S 3.1.1).} (log[r] = 2.199($K-L$) - 2.148) in Figure \ref{fig_duklha}
(\textit{bottom left}).

\subsubsubsection{H$\alpha$ Emission}

Balmer series emission, especially H$\alpha$, is the most common diagnostic
of circumstellar accretion onto young stars.  The large emission-line
fluxes are believed to be generated in the partially optically thin
channeled accretion flow \citep[e.g.,][]{muzerolle98}, although some
chromospheric emission is likely to be present.  In order to account for
the spectral type dependence of the possible underlying chromospheric
H$\alpha$ emission, we define a 'normalized' EW[H$\alpha$] as
nEW[H$\alpha$] $=$ EW[H$\alpha$] - EW[H$\alpha$]$_{w/c}$, where
EW[H$\alpha$]$_{w/c}$ is the equivalent width value used to distinguish
CTTSs from
WTTSs (\S 2.1).  This definition yields a more convenient distinction
between WTTSs and CTTSs since WTTSs have a positive nEW[H$\alpha$] and
CTTSs have a negative nEW[H$\alpha$].  In Figure \ref{fig_duklha}
(\textit{top right}) the nEW[H$\alpha$]s of the single T Tauri stars are
compared to their optical excesses.  All sources with strong H$\alpha$
emission (nEW[H$\alpha$] $<$ 0) have detectable optical excesses, and the
emission and excesses are modestly correlated.  One possible concern
is that larger continuum excesses may lead to smaller EW[H$\alpha$]s since
the EW[H$\alpha$] is simply a ratio of H$\alpha$ to continuum flux.
Calculating the luminosity of H$\alpha$ emission eliminates the continuum
dependence.  In Figure \ref{fig_duklha} (\textit{bottom right}), the
H$\alpha$ luminosities are also compared to their optical excesses for the
same sample of T Tauri stars.  The continuum levels used to determine the
H$\alpha$ luminosities are estimated from reddening corrected $R_c$
magnitudes.  Although the H$\alpha$ luminosities are modestly correlated
with the levels of optical excess emission, there is still considerable
scatter in the relation and there is an observable overlap in the H$\alpha$
luminosities for systems with and without optical excesses.  In the
following analysis, only the nEW[H$\alpha$] is used as a tracer of
circumstellar accretion.

\subsubsection{Assigning T Tauri Types}

The nEW[H$\alpha$]s and the $K-L$ colors show the most distinct
distributions of values for systems with optical excesses compared to
systems without optical excesses (Fig \ref{fig_duklha}). Thus, both
could be used for classifying T Tauri types (CTTS versus WTTS; \S 2.1).  We
adopt the nEW[H$\alpha$] for distinguishing CTTSs and WTTSs since it is
the more commonly used determinant.  High accretion stars, however, are
distinguished from CTTSs with more moderate accretion rates by their red
$K-L$ colors (\S 3.1.1).  The assigned T Tauri types for the components of
the binary sample are listed in Table \ref{tab_bin_stel}.  In the absence
of spatially resolved H$\alpha$ measurements, $K-L$ colors are used to
assign the T Tauri types: $K-L$ $< 0.4$ are WTTSs (V410 Tau A), $0.4 <$
$K-L$ $< 1.4$ are CTTSs (UZ Tau Ba \& Bb).  These types are marked in Table
\ref{tab_bin_stel} with a
colon.  The T Tauri types of the components without spatially resolved
H$\alpha$ or $K-L$ measurements are assigned based on UV excesses (Fig
\ref{fig_duklha}): $\Delta$U $< 0.8$ magnitudes are WTTSs (V773 Tau A \& C,
V410 Tau C), $\Delta$U $> 0.8$ magnitudes are CTTSs (DF Tau A \& B).  These
types are marked with a double colon in Table \ref{tab_bin_stel}.  The
types assigned from $K-L$ colors and $U$-band excesses are consistent with
the T Tauri types assigned to the unresolved systems as determined from
H$\alpha$ emission (Table \ref{tab_sample}).

One single star (IQ Tau) and several binary star components (IS Tau B, LkHa
332/G1 A, LkHa 332/G1 B, FV Tau/c A) have weak H$\alpha$ emission
(i.e. below the CTTS limit), but red $K-L$ colors ($> 0.5$ magnitudes).
Conversely, a few stars (GM Aur, FP Tau, FO Tau B) have strong H$\alpha$
emission, but photospheric $K-L$ colors.  The T Tauri types assigned to
these peculiar stars therefore depends on whether the nEW[H$\alpha$]s
or $K-L$ colors are used; here the nEW[H$\alpha$]s are used.  In the
majority of cases, however (40/43 = 93\% of single stars; 33/38 = 87\% of
binary star components), the T Tauri types assigned by the nEW[H$\alpha$]s
and the $K-L$ colors are consistent.  For the two stars GM Aur and FP Tau
with $K-L$ colors typical of WTTSs, their CTTS type assigned from the
strength of H$\alpha$ emission are supported by their optically veiled
spectra \citep{heg95}.  No optical veiling measurements are available for
the other peculiar stars.

\subsubsection{The Observed Circumstellar Properties of Binary Star
Components}

Mass accretion rates for the components of the binary systems resolved
optically are calculated following the same methodology used for the single
T Tauri stars.  For high accretion stars, these values are very uncertain 
because of the considerable uncertainty in the stellar parameters
(\S 3.1.1).  Accretion rates determined for WTTSs are assumed to be upper
limits; an upper limit of $10^{-10}$ M$_\odot$ yr$^{-1}$ is assigned to
V773 Tau A, which has a small negative $U$-band excess.  The $\Delta U$s
and mass accretion rates are listed in Table \ref{tab_bin_stel}, along with
the $K-L$ colors and EW[H$\alpha$] measurements.

In Figure \ref{fig_pacc3}, the mass accretion rates, the $K-L$ colors, and
the nEW[H$\alpha$]s of primary stars within both CTTS and WTTS systems are
plotted versus the projected separation of their companion.  Systems with a
high accretion star are not included since it is unclear which component is
the true primary (\S 3.1.3).  For comparison, the corresponding accretion
diagnostics of single CTTSs and single WTTSs (Appendix B) are plotted with
arbitrary separations.  For all three accretion diagnostics, the CTTS
primaries have distributions that are indistinguishable from the single
CTTS distributions.  This is true independent of the binary separation.
The presence of a companion star, even as close as $\sim 10$ AU, does not
appear to disrupt the distribution of inner circumstellar material
sufficiently to diminish or enhance the mass accretion rate.  
The majority of WTTS primaries have accretion diagnostics that are
consistent with the single WTTS sample, and indicative of little or no
accretion.

In Figure \ref{fig_psacc3}, the mass accretion rates, the $K-L$ colors, and
the nEW[H$\alpha$]s of secondary stars are compared to their more massive
companions.  Pairs are distinguished based on the T Tauri type of the
system (Table \ref{tab_sample}) and binary separation (close versus wide).
These comparisons suggest that, independent of the accretion
diagnostic used, primary stars generally have accretion signatures that are
comparable to or larger than that of their associated companion, implying a
higher mass accretion rate.  \citet{gws97}, \citet{duchene99b}, and
\citet{pm01} have reported similar results.  We find this to be true for
both close and wide pairs; the distributions of relative accretion
diagnostics (primary versus secondary) for close and wide pairs are not
significantly different.  
The dominance of the primaries' accretion signatures is the most
pronounced in systems with small mass ratios (i.e. systems with the most
discrepant masses).  This is illustrated in Figure \ref{fig_qacc3}, which
plots the differences in the mass accretion rates, the $K-L$ colors, and
the nEW[H$\alpha$]s between primaries and secondaries versus their mass
ratios.  Although the small number of low mass ratio systems limits the
interpretation, a general trend is evident considering all 3 diagnostics.
In comparable mass systems (m$_\textrm{s}$/m$_\textrm{p}$ $>$
0.8), either component can have the dominant accretion signatures.  In
smaller mass ratio systems (m$_\textrm{s}$/m$_\textrm{p}$ $<$ 0.8),
however, primaries systematically have the dominant accretion signatures.
Although pairs with high accretion components are not included in these
comparisons, these pairs would be a strong exception to this result if high
accretion stars are lower mass components.  However, \citet{koresko97}
have shown that many high accretion stars are, bolometrically, the more
luminous component and thus may be the more massive component.

The presence of circumprimary and circumsecondary disks is investigated
using the spatially resolved accretion disk signatures of the binary sample
(Fig \ref{fig_psacc3}).  As discussed in \S 3.2.2, both the nEW[H$\alpha$]
and the $K-L$ color have relatively clean
breaks between accreting (CTTS) and non-accreting (WTTS) stars.  We
therefore identify circumstellar disks based on nEW[H$\alpha$]s less than
zero or $K-L$ colors greater than 0.4 magnitudes.  As shown in Figure
\ref{fig_psacc3}, systems which support a circumsecondary disk, but no
circumprimary disk are rare.  Only 1 of 35 systems with $K-L$ measurements
and 0 of 16 systems with nEW[H$\alpha$] measurements show this
configuration.  In contrast, there are many systems (7 of 35 based on $K-L$
and 5 of 16 based on nEW[H$\alpha$]) with a circumprimary disk but no
circumsecondary disk.  Thus, if only one circumstellar disk exists, it is
almost always associated with the more massive star.  The existence of a
component's disk, however, appears to be correlated with the existence of
its companion's disk for close systems.  In this more general comparison in
which mass is not relevant, high accretion stars are included as CTTSs.
The correlation of disk existence, or equivalently of T Tauri type, holds out
to a separation of $\sim 210$ AU.  Of the 16 pairs with separations less
than 210 AU and with spatially separated nEW[H$\alpha$] measurements (2 WTTS
and 14 CTTS systems), only three systems are mixed types (FV Tau/c, IS Tau,
V807 Tau).  Of the 30 pairs with separations less than 210 AU and with
spatially separated $K-L$ measurements (21 CTTS and 9 WTTS systems), only 4
are mixed types (FO Tau, FQ Tau, VY Tau, V807 Tau).  Assuming a fixed
distribution of primaries (WTTS or CTTS) and an equal probability of having
either a WTTS or a CTTS companion \citep{hartmann91}, the probability to
get the observed number of mixed systems from each of the two samples is
$6\times10^{-4}$ (C$^{3}_{16}(\onehalf)^{16}$) and $3\times10^{-5}$
(C$^{4}_{30}(\onehalf)^{30}$), respectively.  These low probabilities
strongly suggest that the relative T Tauri types are correlated.
In contrast, of the 8 pairs with separations greater than 210 AU and with
spatially separated nEW[H$\alpha$] measurements (2 WTTS and 6 CTTS systems),
5 are mixed types.  Similarly, of the 11 wider pairs with spatially
separated $K-L$ measurements (10 CTTSs and 1 WTTSs), 4 are mixed types.  The
circumstellar disks in systems wider than 210 AU are more consistent with
random pairing.

\section{Discussion}

\subsection{Circumstellar Disks in Binary Star Systems}

Millimeter observations of binary T Tauri stars suggest that the
circumprimary and circumsecondary disk masses of close binary ($< 100$ AU)
systems may be diminished relative to the disk masses of single T Tauri
stars \citep{ob95, dutrey96, jensen96}.  The diminished masses are
attributed to disk truncation by the companion star.  Since it is
statistically unlikely that these binary stars have disks that are just on
the verge of complete depletion, the disk mass estimates and the disk
accretion rates can be used to calculate a characteristic disk lifetime
(M$_{\textrm{\scriptsize disk}}$/$\dot{\textrm M}$).  Following this
simple procedure, several binary star studies have found that the disk
lifetimes for binary stars, calculated from the best available accretion
rates and age estimates, are typically less than the ages of the stars
\citep{mathieu94, ps97, duchene99b}.  Thus, it is somewhat surprising that
these binary stars are still actively accreting.  However, \citet{ps97}
found that single stars typically have disk lifetimes that are also less
than their ages, which may imply systematic errors in their assumed stellar
ages and/or circumstellar properties.  The new mass accretion
rates and stellar ages calculated self-consistently here for single stars
and for binary star components using spatially resolved measurements will
allow us to more carefully investigate this potential timescale problem.

In the left-hand panel of Figure \ref{fig_disklife}, the measured mass
accretion rates are combined with disk masses from the literature to
calculate the disk lifetimes, which are then compared to the stellar
ages for both single and binary T Tauri stars.  Disk masses are taken from
\citet{ob95}, \citet{beckwith90}, \citet{dutrey96} for UZ Tau A and B, and
\citet{gds99} for GG Tau A\footnote{The circumstellar disk mass estimate
for GG Tau A is based on measurements which spatially resolve the
circumstellar disk emission from the circumbinary emission \citep{gds99}.}.
Although these millimeter studies do not spatially resolve circumprimary
and circumsecondary disks, the disk mass estimates for each system should
nevertheless be a reasonable approximation of the combined circumstellar
disk mass.  The mass accretion rates for the binary systems are the sums
determined for all components and the ages are the averages of all
components.  Systems with a high accretion star are not included because of
the uncertainty in their ages and mass accretion rates.  As this figure
illustrates, the single stars are distributed evenly about the line that
represents disk lifetimes equal to stellar ages.  In contrast, the binaries
are systematically below this line, implying that the binary star disk
lifetimes are significantly shorter than their stellar ages.  For the
binary stars, this effect shows no significant dependence on separation,
although only 2 of the 12 binary stars have separations larger than 100 AU.

To quantify the comparison of disk lifetimes and stellar ages for single
and binary T Tauri stars, the right-hand panel of Figure \ref{fig_disklife}
shows histograms of the logarithm ratios of stellar ages to disk lifetimes.
Limits are treated as actual values; single stars have only a slightly
smaller fraction of disk lifetime estimates that are based on disk
mass upper limits than binary stars (36\% for singles, 42\% for binaries).
Single stars have disk lifetimes that are, on average, comparable to their
ages (mean log[Age/Disk Lifetime] $= -0.04 \pm 0.18$ dex), while binary
stars have disk lifetimes that are, on average, less than their ages by
nearly a factor of 10 (mean log[Age/Disk Lifetime] = $0.95 \pm 0.34$
dex; a 2.8$\sigma$ difference from zero).  However, a direct comparison of
the log[Age/Disk Lifetime] distributions for singles and binaries using a
K-S test implies that they differ only at the 80\% confidence level, a
1.2$\sigma$ difference (Figure \ref{fig_disklife}).  The discrepancy
between the disk lifetimes and the ages of binary stars, as well as
discrepancy between the distributions of log[Age / Disk Lifetime] for
binaries and singles increase substantially, though, if high accretion
stars are included by assuming that the mass accretion rates determined for
DG Tau and DR Tau from $IUE$ spectra are typical for high accretion stars
\citep[\S 4.3;\,][]{gullbring00} and that they have their companions' or an
average age.  In this case, the mean log[Age/Disk Lifetime] for binaries
increases to $1.47 \pm 0.37$ dex, while for singles it remains at $0.11 \pm
0.18$ dex, and the log[Age/Disk Lifetime] distributions differ at the 99\%
level according to a K-S test.  Thus, although the properties of high
accretion stars are uncertain, their properties support the trends
identified for other T Tauri stars.  We conclude that while single T Tauri
stars have disk lifetimes comparable to their ages, close binary T Tauri
stars have disk lifetimes that are roughly 1/10 of their ages.  

One solution to this timescale problem is that the
circumstellar disks are being replenished from circumbinary reservoirs.  In
support of this, the close binaries GG Tau and UY Aur are known to be
surrounded by substantial circumbinary material \citep{dgs94, roddier96,
gds99, close98, duvert98}, while other systems show spatially extended
circumbinary nebulosity \citep[e.g., FS Tau, T Tau, CoKu Tau 1; \S 2.3.2,
][]{stapelfeldt98a, padgett99}.  The majority of these reservoirs, however,
are not likely to be very massive since the millimeter upper limits of many
of the close binaries correspond to circumbinary mass upper limits of $\sim
0.005$ M$_\odot$, assuming optically thin emission \citep{jensen96}.
Nevertheless, the addition of mass in this amount to the circumstellar
disks of close binaries would be enough to bring the disk lifetimes to
within 2$\sigma$ of the ages, and possibly reconcile the discrepancy.  We
therefore consider the possibility that some circumbinary material is
present and is replenishing the circumstellar disks.  Numerical simulations
have demonstrated how this may occur \citep{al96, bb97}.  For example,
replenishment from a circumbinary disk can successfully explain the
episodic accretion characteristics of DQ Tau, a 16 day spectroscopic binary
in which replenishment from a circumbinary reservoir must be occuring in
order to sustain the observed accretion rates \citep{mathieu97, basri97}.
Although the binaries studied here have orbital periods which are too long
to observe such episodic effects, the consequence of replenishment in these
systems may explain their long disk lifetimes.

The flow of circumbinary material onto the individual circumstellar disks
depends to a large extent on its angular momentum relative to the binary
star.  If the circumbinary material has sufficient angular momentum to form
a circumbinary disk, then the lower mass object is expected to be
preferentially replenished \citep{al94}; this component has closer
encounters with the reservoir of circumbinary material.  Alternatively, if
the angular momentum of the surrounding material is insufficient to form a
circumbinary disk, then the in-falling material is expected to
preferentially replenish the circumprimary disk; this material will fall
toward the center of mass \citep{bb92, bb97}.  Thus if replenishment is
occuring for close (10 - 100 AU) binaries as suggested by their delayed
depletion, the angular momentum of this material may be constrained by
studying the relative circumstellar properties in binary systems.

The result that there are many systems with only circumprimary disks, but
that systems with only circumsecondary disks are very rare, implies that
circumprimary disks are longer lived.  This finding is in contrast to the
predictions of the high angular momentum, circumbinary disk replenishment
scenario.  This scenario preferentially replenishes the circumsecondary's
disk and this, in conjunction with the secondary's lower accretion rate due
to its lower mass (\S 4.2), should result in the secondary having a longer
disk lifetime.  The observations are in better agreement with the low angular
momentum replenishment scenario.  In this case, the circumprimary's disk
will be preferentially replenished, allowing it to survive longer than
the circumsecondary's disk.  The higher accretion rates of primary stars
(Fig \ref{fig_psacc3}) further strengthens the case for preferential primary
replenishment, since otherwise they would deplete their disk first,
assuming equal disk masses.  We note, however, that \citet{armitage99} have
proposed an alternative explanation for the longer disk lifetimes of
primary stars in which neither disk is replenished.  The more massive
component within a binary system is capable of supporting a more massive
circumstellar disk due to its larger Roche lobe.  This disk will
consequently survive longer than the circumsecondary disk, presuming
similar accretion rates.  Replenishment scenarios are nevertheless a likely
possibility for close binary T Tauri stars given their apparently short
disk dissipation timescales and the direct evidence of circumbinary
material in some systems.

The relative T Tauri types and binary mass ratios provide additional
support for replenishment from a circumbinary reservoir.  As suggested by
\citet{ps97}, replenishment from a common reservoir (e.g., circumbinary)
may co-regulate the evolution of the binary star components
\citep[e.g.,][]{bb97} and may result in circumprimary and circumsecondary
disk dissipation timescales that are similar.  Both components would be
replenished, with perhaps a modest preference for the primary as noted
above, until the circumbinary reservoir is depleted.  In such a scenario,
CTTS primaries should preferentially have CTTS secondaries.  This
prediction is supported by the relative T Tauri types of binaries with
separations less than 210 AU, which are
strongly correlated (\S 3.2.3).  \citet{ps97} and \citet{duchene99b} have
found a similar correlation of T Tauri type.  Although they suggest this
correlation may extend to wider separations than 200 AU, their
samples of wider binaries are small and, in the latter study, biased by
stars that are not currently believed to be of the T Tauri class
\citep[e.g. HBC 352, HBC 355, HBC 360;][]{martin94}.  The mixed pairing of
wider systems studied here suggests a maximum binary separation for this
correlation.  Pairs with a separation of up to $\sim$ 200 AU appear to
be regulated by the same circumbinary reservoir, while wider pairs are not.  

This separation limit is only
slightly larger than the separation at which a slight break occurs in the
distribution of mass ratios ($\sim 100$ AU; Fig \ref{fig_hst6_mass}).
Accretion from a circumbinary reservoir can also
explain this.  Under the assumption of formation by fragmentation (\S 4.4),
only a small fraction ($\lesssim 10$\%) of the final stellar masses are
contained in the initial protobinary fragments for systems of separation
less than $\sim 100$ AU \citep{boss88, bate00}.  For these systems the
majority of the mass of the stellar components is accreted from
circumbinary material \citep{bb97}.  Consequently, \citet{bate00} predict
that close binary systems are more likely to have high mass ratios than
wide binary systems; the closer the components are to the center of mass of
the system, the more equally they will receive accreted mass.  This effect
is not expected to yield a strong mass ratio dependence on separation over
the range of separations explored here (10 - 1000 AU), and thus is
consistent with the weak trend seen in Figure \ref{fig_hst6_mass}.

\subsection{Evidence for a Mass Dependent Accretion Rate}

As Figures \ref{fig_psacc3} and \ref{fig_qacc3} illustrate, primaries
generally have dominant accretion signatures, and their dominance is the 
most significant in systems with small mass ratios.  Since the presence
of a close companion appears to have little effect on the mass accretion
rate (Fig \ref{fig_pacc3}), this suggests that the higher relative
accretion rates of primary stars may simply be a consequence of their
higher relative mass.  To investigate this, in Figure \ref{fig_macc3}
the mass accretion rates, $K-L$ excesses\footnote{A $K-L$ excess is
used here in order to eliminate the small spectral type dependence of the
underlying
photosphere, since this information is available for all stars with mass
estimates.  The $K-L$ excess is defined as the difference between the
observed $K-L$ color and the reddened photospheric $K-L$ color (\S 3.1.1).}, 
and nEW[H$\alpha$]s of CTTS binary components are plotted
versus their stellar mass.  Only components with masses derived from
optical measurements are shown; high accretion stars are excluded.
Primary and secondary stars are distinguished, but close and wide
pairs are not since both sets have similar distributions of mass accretion
rates, $K-L$ colors, and nEW[H$\alpha$]s (Fig \ref{fig_psacc3}).  The
accretion diagnostics of the single CTTS sample are also included.

The mass accretion rates shown in Figure \ref{fig_macc3} show a general
trend of decreasing mass accretion rate with decreasing stellar mass.
The Spearman rank correlation coefficient between log[$\dot{\textrm{M}}$]
and log[mass] is 0.57, with a probability of $< 0.1$\% of being drawn from a
random distribution.  Nevertheless, stars near the median mass ($\sim 0.7$
M$_\odot$) experience mass accretion rates which span 2.5 orders of
magnitude.  Thus, although the mass accretion rate appears to be correlated
with stellar mass, it is not a one-to-one relation.  The $K-L$ color excesses
and nEW[H$\alpha$]s also plotted in Figure \ref{fig_macc3} are not as well
correlated with stellar mass.  Although the $K-L$ excesses and the
nEW[H$\alpha$] measurements for stars less massive than 0.3 M$_\odot$
(log[M] $<$ 0.5) are below the average value for higher mass stars, the
small number of low mass T Tauri stars with accretion diagnostics limits
the statistical significance of this comparison.  For all 3 accretion
diagnostics however, the distribution of primaries, secondaries, and
single stars with similar masses are indistinguishable.  The dominant
mass accretion signatures of primary stars relative to their companions
is therefore most likely because of their higher
relative mass, and not a consequence of being in a binary system.

\subsection{High Accretion T Tauri Stars}

Of the 82 binary star components in Table \ref{tab_sample} with spatially
resolved NIR measurements, 9 (11\%) are classified as high accretion stars
based on their very red NIR colors (\S
3.1.1).  This fraction is similar to previous estimates.  \citet{zw92}
estimate that approximately 5\% of binary star components (within 10\% of
binary systems) have similar red colors, although in these previous studies
they have usually been referred to as 'infrared companions' \citep[see
also][]{ghez97}.  We
have not adopted this terminology here, however, since the word
'infrared' has come to imply that these sources are optically invisible,
which is often not the case, and the word 'companion' implies that this
phenomenon occurs only within binary star systems, although there is no
convincing evidence to support this.  Of the 52 single stars in Table
\ref{tab_sing}, 3 (6\%) are classified as high accretion stars.  Although
this fraction is slightly less than the fraction in the binary sample
studied here, there
is an observational bias toward finding these optically faint stars near a
known T Tauri star because of directed companion searches at NIR
wavelengths \citep[e.g.,][]{gnm93, leinert93, simon95}.  This bias may also
explain why the known single high accretion stars are the brightest members
of this class; isolated high accretion stars would typically be too faint,
for their optical colors, to be distinguished from the background
population in optical photometric surveys
\citep[e.g.,][]{briceno98}.  Deep NIR surveys, ideally including $L$-band
measurements, offer the best hope of identifying these
peculiar young stars \citep[e.g.,][]{haisch00, lada00}.

The similarities in the spectral energy distributions of Class I sources,
which are believed to be at an earlier stage of evolution than T Tauri stars
\citep{lw84}, and some high accretion stars like T Tau B have led many
astronomers to speculate that some high accretion companions are also
protostars at an earlier stage of evolution than the T Tauri stage 
\citep[e.g.,][]{dsz82}.  This scenario, however, is inconsistent with the
coeval ages of binary stars presented in this study, and the favored binary
star formation mechanism that produces them (\S 4.4).  It is more likely
that the high accretion stars are relatively normal T Tauri stars
experiencing unusually high accretion.  For example, \citet{gullbring00},
using \textit{IUE} archival spectra, estimate the mass accretion rates for
the single high accretion stars DG Tau and DR Tau to be 3-5 $\times$
$10^{-7}$ M$_\odot$yr$^{-1}$, which is nearly a factor of 100 (2.9$\sigma$
in log[$\dot{\textrm M}$]) above the mean accretion rates of CTTSs (\S 3.2.1).
The similar properties of Class I sources thus likely stem from their
similar high levels of accretion that are above the norm for T Tauri stars
\citep{kenyon98}.  In several cases, the optical spectra of high accretion
stars strikingly resemble those of Class I sources \citep[e.g., HV Tau C
versus IRAS
04264+2433;][]{mm94, kenyon98}, especially the prominence and strength of
forbidden line-emission.  The red $K-L$ colors of both high
accretion stars and Class I sources demonstrate that it is not possible to
unambiguously identify Class I sources based solely on $K-L$ colors as is
sometimes done \citep[e.g.,][]{haisch00, lada00}.  Approximately
10\% of a T Tauri population may be high accretion stars.  If the enhanced
accretion of these stars is an episodic phenomenon \citep{ghez91,
koresko97}, this fraction of high accretion stars would then suggest that T
Tauri stars spend approximately 10\% of their T Tauri life in this enhanced
accretion phase.

With such large levels of excess emission, one would expect high accretion
stars to appear over-luminous relative to the majority of T Tauri stars.
Surprisingly, the optical measurements reveal the opposite.  Figure
\ref{fig_bpsbin} illustrates that high accretion stars are, on average,
systematically under-luminous relative to other T Tauri stars of similar
temperature and/or color.  The most likely explanation for these low
luminosity estimates is that the extinction estimates are too low.  There
are two reasons to expect this.  First, the substantial UV excesses
associated with high accretion rate stars may extend considerably into the
$V$-band, giving the systems a bluer, less extincted appearance.  Second,
several high accretion stars display considerable extended emission
\citep[e.g., HL Tau, HV Tau, FS Tau;][]{stapelfeldt95, mb00, krist98} and
in some
cases are highly polarized sources \citep[DG Tau, HL Tau;][]{bastien82}.
If much of the optical flux from high accretion stars is attributable to
scattered light, the preferential blue scattering from small dust grains
will again give the system a bluer and less extincted appearance.  We
emphasize, however, that the high accretion star phenomenon can not be
explained simply as an edge-on disk orientation effect.  For example, the
spatially resolved edge-on disk system HK Tau B \citep{stapelfeldt98b,
koresko98} is a borderline CTTS/WTTS (nEW[H$\alpha$] = +2.5 \AA;
Monin et al. 1998) with a modestly reddened photospheric $K-L$ color (0.46
magnitudes; Table \ref{tab_nir}).  Nevertheless, the properties of high
accretion stars imply that high levels of accretion are correlated with
high levels of extinction and scattered light emission.  In light of this
correlation, we consider the optically faint CTTSs FW Tau C (Table
\ref{tab_bin_stel}), LkHa 358 (Table \ref{tab_sing}), and ITG 33a
\citep{itg96, martin00}, which have no $L$-band measurements, to be
candidate high accretion stars.

\subsection{Implications for Binary Star Formation}

As summarized by \citet{clarke95}, the proposed binary star formation
scenarios typically fall into three main categories: disk instabilities,
capture, and core fragmentation.  Each of these scenarios produces binary
star populations with distinct stellar and circumstellar properties that
can in principle be tested, and this has been the goal of many
young binary star studies \citep{ck79, hss94, koresko95, bz97, gws97,
wlk01}.  The combined results of these studies favor fragmentation
\citep[e.g.,][]{boss88} as the dominant binary star formation mechanism.  
As summarized in \citet{gws97}, the observations that binary star
components are coeval, that secondary star mass and mass ratio
distributions are independent of the primary star mass, that secondary star
masses are typically much greater than T Tauri star disk masses, and that
GG Tau A retains a well defined
circumbinary disk, all uphold a model in which the systems
form via fragmentation.  The properties of the larger sample of systems
with resolved optical measurements presented here strengthen the arguments
for the fragmentation scenario.  In particular, the relative ages of binary
star components in Taurus-Auriga are shown to be more similar than the
relative ages of randomly paired single T Tauri stars in Taurus-Auriga.
Although the observed coevality of these components ($\lesssim$ 1 Myrs) is
still considerably larger than typical collapse timescales ($\sim$ 0.1
Myrs), this is the first statistically significant evidence that
demonstrates that the components of binary stars are related in a way that
single stars are not.
Secondly, the survival of circumstellar disks in close binary systems, the
correlation of T Tauri types, and the tentative suggestion of a mass ratio
dependence on separation, all provide additional support for circumbinary
structures.  Capture encounters are thought to preclude the formation of
stable circumbinary material \citep{hall96}, but these structures are
believed to be a natural consequence of core fragmentation.
Altogether, the stellar and circumstellar properties of binary stars
strongly support the conclusion that fragmentation, as opposed to disk
instabilities or capture, is the dominant binary star formation mechanism.

\subsection{The Overabundance of T Tauri Companions}

The secondary star mass estimates can be used to eliminate one proposed
explanation for the overabundance of T Tauri star companions.  Over the
separation range considered here (10 - 1000 AU), T Tauri stars are 2-3
times more likely to have a companion star than solar-type main-sequence
stars \citep{ghez95a}.  One suggested explanation for this is that the
surveys of young stars are more sensitive to low mass companions and
therefore detect low mass stars (or possibly substellar objects) that are
missed in surveys of older main-sequence stars.  Since most main-sequence
surveys have detection limits near the stellar/substellar boundary
\citep[0.08 M$_\odot$;][]{dm91}, this explanation would require
that nearly half of T Tauri companions have masses at or below the
substellar limit.  In contrast to this, all of the companion masses
determined here except for GG Tau Bb (1/21) are above the substellar limit.
Thus the overabundance of T Tauri companions cannot be explained by a
substantially higher fraction of very low mass companions.
The high binary fraction of T Tauri stars may instead be a
consequence of the star formation process in low density T Associations.  
The star formation process in giant molecular clouds (e.g., Orion),
which is believed to be the birthplace for the majority of solar-type
stars \citep{ms78}, appears to produce a binary fraction than that
is consistent with solar-type main-sequence stars \citep{petr98,
duchene99a}.

\section{Summary and Conclusions}

We have carried out an extensive high spatial resolution survey of multiple
T Tauri star systems in Taurus-Auriga.  Three sets of \textit{new}
observational data are presented: (1) high resolution speckle and direct
imaging measurements at $K$ and $L$ of nearly all binary systems with
separations of 0\farcs14 - 7\farcs0 (20 - 1000 AU), (2) high resolution
space-based optical, UV, and narrow-band H$\alpha$ observations of 11
binaries with separations of 0\farcs07 - 0\farcs7 (10 - 100
AU), and (3) spatially separated, moderate resolution (R $\sim 1500$)
optical spectra of four close binaries (GG Tau, GH Tau, V955 Tau,
and XZ Tau; separations 35 - 50 AU).  In addition to the previously
known components, these observations revealed a new companion in the
FW Tau system (separation $=$ 2\farcs29), making this a triple star system.
The new observations presented here, in combination with previous
measurements, are used to determine the stellar and circumstellar
properties of the components of binary stars in Taurus-Auriga.  A
carefully selected sample of single T Tauri stars in this region is also
identified for comparison (Appendix B).  

Masses and ages for the components with optical measurements are determined
from comparisons with PMS evolutionary models.  The ages derived for the
binary star components range from $< 1$ Myrs to 19 Myrs and the stellar
masses range from 0.042 M$_\odot$ to 2.11 M$_\odot$.  The distributions of
these ages and masses are indistinguishable from the distributions of
single T Tauri stars in Taurus-Auriga.  Except for GG Tau Bb, the secondary
star masses are all above the substellar limit, and this result eliminates
one proposed explanation for the high frequency of companions to T Tauri
stars relative to main-sequence stars.  The overabundance of T Tauri
companions cannot be explained by a substantially higher fraction of very
low mass companions that would have been missed in main-sequence surveys.

The circumstellar accretion disks within binary star systems are studied in
this work through three diagnostics - the infrared color $K-L$, UV excess,
and H$\alpha$ emission.  The strength of these accretion signatures for
primary stars are similar to single stars, which suggests that companions
as close as 10 AU have
little effect on the mass accretion rate.  The inferred accretion rates
are used in combination with disk mass estimates from the literature to
compare the disk lifetimes and the stellar ages for both single and binary
T Tauri stars.  Although single stars have disk lifetimes comparable to
their ages, binary stars have disk lifetimes that are roughly 1/10
of their ages.  Thus, the continuing disk accretion in binary systems
suggests that the supposedly truncated circumprimary and circumsecondary
disks are being replenished.

The circumstellar properties are used to compare circumprimary and
circumsecondary disks.  While there are several systems that contain only
a circumprimary disk, systems that contain only a circumsecondary disk are
rare.  This suggests that, on average, circumprimary disks survive longer
than circumsecondary disks.  Primary stars also appear to accrete at a
higher rate than their lower mass companions.  The longer disk lifetime
of circumprimary disks, despite their higher accretion rates, suggests that
circumprimary disks are being preferentially replenished, possibly from a
circumbinary reservoir that has low angular momentum relative to the
binary.  This replenishment scenario is further supported by the high mass
ratios of close pairs ($\lesssim$ 100 AU) as well as the similar T Tauri
types of binaries with separations of less than $\sim$ 200 AU.
\citet{bate00} predict that the majority of the final stellar mass for the
closest systems is accreted from circumbinary material and that the closest
systems will accrete more equally than wider systems.  Consequently, the
closest pairs are driven toward mass ratios of unity and have components
with comparable disk lifetimes, consistent with the observations.

The higher mass accretion rates of primary stars relative to secondary
stars are most likely due to their larger masses.  Primary, secondary, and
single stars of similar mass have similar mass accretion rates, but
higher mass stars generally have larger accretion rates than lower mass
stars.  Approximately 10\% of T Tauri stars, including both single stars
and binary star components, can be distinguished as high accretion stars
by their uncharacteristically red near- and mid-infrared colors.
Optically, high accretion stars are, on average, distinctively faint and
exhibit heavily veiled spectra with unusually strong forbidden-line
emission; high extinction and scattered light emission appear to be
correlated with enhanced accretion.  High accretion stars are probably not
at an earlier stage of evolution as has been proposed.  The similarities
they share with younger protostars at optical and infrared wavelengths stem
from their similar accretion rates, which are above the norm for T Tauri
stars, as opposed to their ages.

Both the stellar and circumstellar properties of this large binary star
sample favor fragmentation \citep[e.g.,][]{boss88} as the dominant binary
star formation process.  The relative ages of binary stars, which are
currently limited by measurement uncertainties, imply that binary star
components are coeval to $\sim$ 1 Myrs.  These relative ages, however, still
only offer weak constraints on the details of the formation since they are
considerably larger than typical collapse timescales ($\sim$ 0.1 Myrs).
Nevertheless, binary stars are shown to be more coeval than randomly paired
single T Tauri stars within the same star forming region, which implies
that they are related in a way that other stars within the same region are
not (i.e. through formation).  Additionally, the case for formation via
fragmentation is strengthened by the direct detection of circumbinary
structures (e.g. GG Tau), and the indirect evidence for circumstellar disk
replenishment from a circumbinary reservoir for close pairs (e.g. short
disk lifetimes, correlations in T Tauri type, high mass ratios).
Circumbinary disks and/or envelopes are a natural outcome of the
fragmentation process, but are difficult to retain in alternative formation
scenarios such as capture.  Overall, the stellar and circumstellar
properties of binary T Tauri star components offer many powerful
constraints on the formation and evolution of binary stars and their
associated circumstellar material.

\acknowledgments

Support for this work was provided by the Packard Foundation and NASA
through grants NAGW-4770 NAG5-6975 under the Origins of Solar Systems
Program and grant numbers G0-06014.01-94A and G0-06735.01-95A from the
Space Telescope Science Institute, which is operated by AURA, Inc., under
NASA contract NAS5-26555.  The authors are grateful to G. Duch\^ene,
J. Patience, L. Prato, and R. Webb for helpful comments and discussions and
appreciate the assistance provided by the FOS instrument scientists
T. Keyes and E. Smith.  In addition, an anonymous referee carefully
reviewed this paper and provided much appreciated comments and suggestions.

\appendix

\section{HST Broad-Band Filter Transformations}

In order to compare the HST broad-band measurements in the Vega magnitude
system with standard photospheric colors and previous ground-based
observations, the broad-band measurements are transformed to the more
standard Johnson-Cousins filter system.  These transformations, which are a
function of spectral type, are calculated with STScI's SYNPHOT package
within IRAF using the dwarf spectral standards of \citet{gs83}.  These
transformations are listed in Table \ref{tab_filt} for spectral types G8
through M8.  Table \ref{tab_filt} also lists the transformation values to
change F675W magnitudes to F656N (narrow-band H$\alpha$) magnitudes in the
Vega magnitude system.  These transformation are derived in the same way
that the broad-band transformations are derived, and are used to estimate
the photospheric contribution to the narrow-band H$\alpha$ flux (see \S
2.3.2).  The last column in Table \ref{tab_filt} lists the ID numbers from
\citet{gs83} of the stars used to establish the transformations.  If
multiple stars are listed, the values are their average; if no stars are
listed, the values are interpolated with respect to temperature from hotter
and cooler spectral types.  Uncertainties are estimated to be 0.02
magnitudes for $U$-F336W, $B$-F439W, and F656N-F675W transformations and
0.01 magnitudes for $V$-F555W, $R_c$-F675W, and $I_c$-F814W transformations
for spectral types earlier than M5.  These uncertainty estimates are based
on variations of the transformation values for stars with similar or the
same spectral type.  Because of insufficient spectral standards, more
conservative values of 0.05 and 0.02 magnitudes, respectively, should be
adopted for the uncertainty in the transformation of cooler spectral types.

The F336W filter has a red leak, which for unextincted K and M dwarfs is
typically 10\% of the observed flux \citep{holtzmann95b}.  However, by
transforming the flux measurements to the Johnson-Cousins system using
transformation derived from the filter throughput curves, the read leak is
accounted for.

\section{Single T Tauri Stars in Taurus-Auriga}

In order to understand the effects of a close companion star and the binary
star formation process, it is useful to have a comparison sample of single
stars.  Table \ref{tab_sing} lists single T Tauri stars in Taurus-Auriga
with spectral types later than K0, with both $V$ and $I_c$
photometry, and with strong Li I \citep[6708 \AA;][]{martin98} absorption.
These single stars were selected from the high resolution imaging studies
of \citet{gnm93, gsw01}, \citet{leinert93}, \citet{simon95},
\citet{sblt98}, and \citet{bra99}\footnote{Although stars have been
detected close to GK Tau and HO Tau \citep{hss94}, optical photometry
demonstrate that they are unlikely to be associated and too faint to bias
the inferred properties.}.  Spectroscopic binaries and radial velocity
variables were also excluded when radial velocity measurements were
available \citep{hartmann86, hartmann87, walter88, mathieu97}.  We
also report that CoKu Tau 4, FN Tau, Haro 6-13, and JH 108 are single
stars \citep{gnm93, leinert93, simon95} that have insufficient optical
photometry to be included \citep{kh95}.  Although all stars in Table
\ref{tab_sing} have no known companions within 10\farcs0, some small
fraction of these 'single' stars may nevertheless have a companion at a
separation less than the survey resolution (typically 0\farcs1) or with a
brightness below the detection limit (typically 2-3 magnitudes fainter than
the target star at 2.2 $\mu$m).  Additionally, near several stars listed
here as single are additional T Tauri stars at separations just above
10\farcs0 (DH Tau \& DI Tau; GI Tau \& GK Tau; \citet{hss94}).  These wide
pairs may be associated.

The HST transformed, Johnson-Cousins photometry for the two single T Tauri
stars imaged in this program, FY Tau and FZ Tau, are listed here for
completeness.  FY Tau: $U = 17.37 \pm 0.04$, $B = 17.11 \pm 0.05$, $V =
15.18 \pm 0.02$, $R_c = 13.67 \pm 0.03$, $I_c = 12.31 \pm 0.02$.  FZ
Tau: $U = 15.98 \pm 0.02$, $B = 16.40 \pm 0.06$, $V = 15.05 \pm 0.07$,
$R_c = 13.60 \pm 0.08$, $I_c = 12.26 \pm 0.04$.  The star near FW Tau,
at a separation of 12\farcs20 and position angle of 246.8$^\circ$ from the
primary, exhibits H$\alpha$ absorption (EW[H$\alpha$] = $+6.5 \pm 0.6$) and
is assumed to be a background star.

The stellar and circumstellar properties of the single stars are derived
following the methodology outlined in \S 3.  These values are listed Table
\ref{tab_sing} along with the optical photometry and spectral type
references.  The values inferred for the high accretion stars should be
used with caution (\S 3.1.1).  The single star sample is plotted on an
H-R diagram in Figure \ref{fig_bpsbin}.  The inferred ages range from $<
1$ Myrs to 19 Myrs and the inferred masses range from 0.087 M$_\odot$ to
1.24 M$_\odot$ (Table \ref{tab_sing}).  The WTTSs and CTTSs have similar
distributions of masses and ages.  If stars with an age upper limit are
assigned an age of 1 Myrs, the mean log[age] of the 29 CTTSs is 6.46
($\sigma = 0.40$) and the mean log[age] of the 20 WTTSs is 6.29 ($\sigma =
0.35$).  The combined sample of 49 stars has a mean log[age] of 6.39
($\sigma = 0.39$).

\clearpage

\begin{figure}
\epsscale{1.0}
\plotone{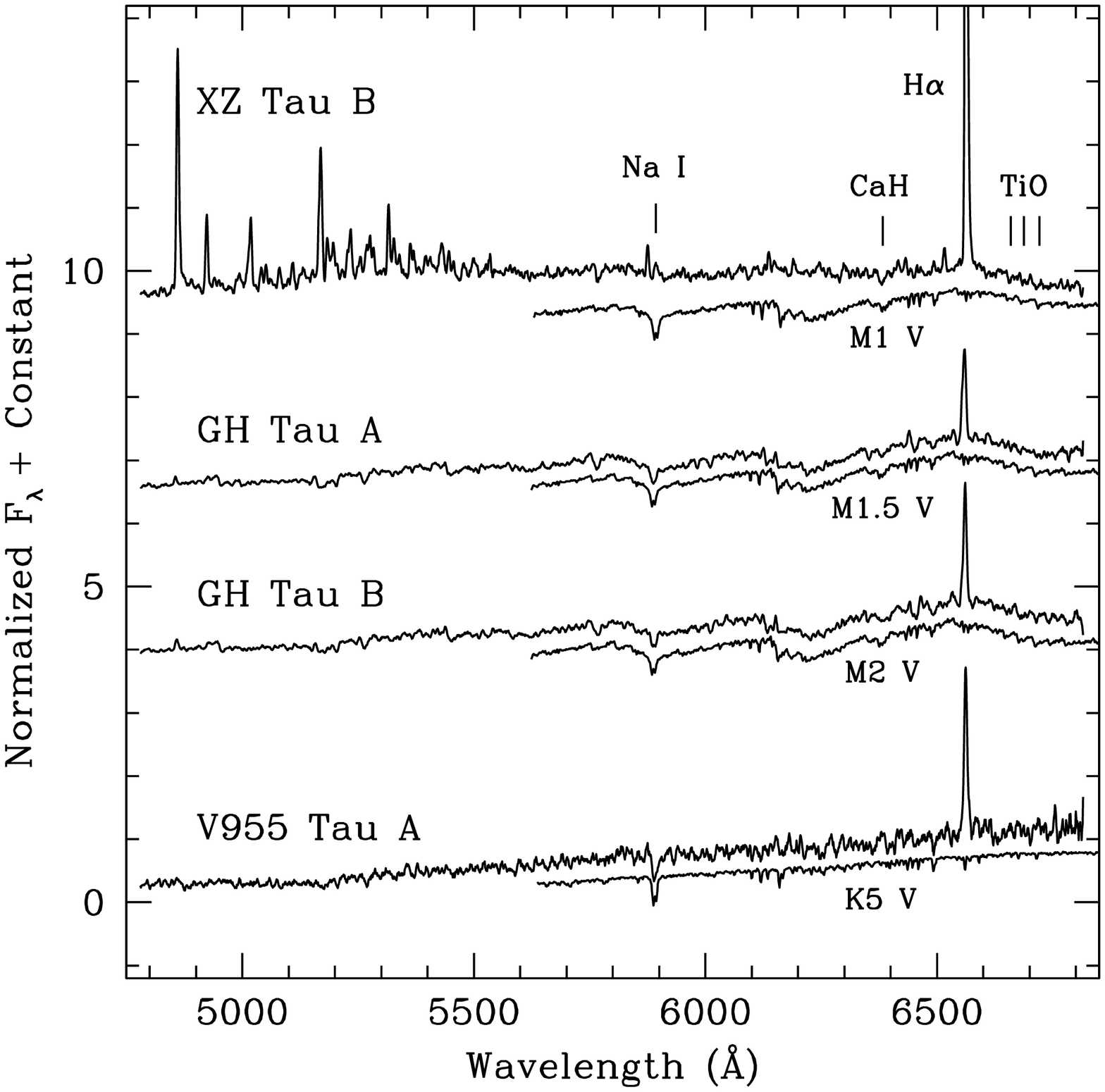}
\caption{Hubble Space Telescope / Faint Object Spectrograph spectra
of XZ Tau B, GH Tau A, GH Tau B, and V955 Tau A, normalized at 6500 \AA.  
The spectra have been offset vertically for clarity, but have the same
relative scaling.  Dwarf spectra of similar spectral type are also
shown, displaced vertically, over the wavelength region used for
spectral classification.  The XZ Tau B spectrum is heavily veiled, but
nevertheless shows faint TiO absorption near 6700 \AA.
\label{fig_fos}}
\end{figure} 

\begin{figure}
\epsscale{1.0}
\plotone{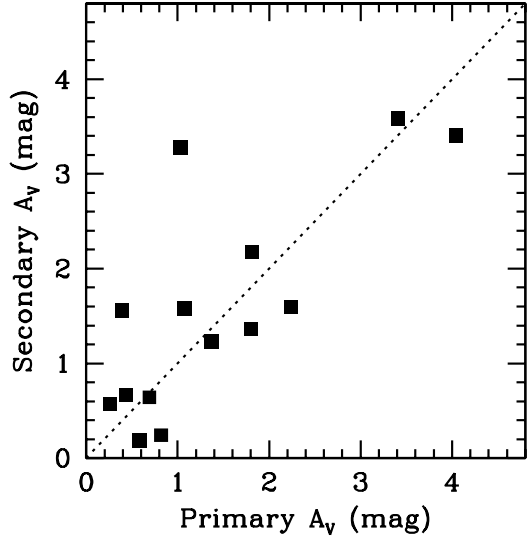}
\caption{The visual extinctions of secondary stars are plotted versus the
visual extinctions of primary stars for binary systems in which the
extinction of each component is determined independently.  Six pairs
are from Taurus-Auriga (Table \ref{tab_bin_stel}); eight pairs are from
Chamaeleon, Lupus, and Ophiuchus \citep{bz97}.  The components of a binary
system typically have similar visual extinctions. \label{fig_bin_av}}
\end{figure} 

\begin{figure}
\epsscale{1.0}
\plotone{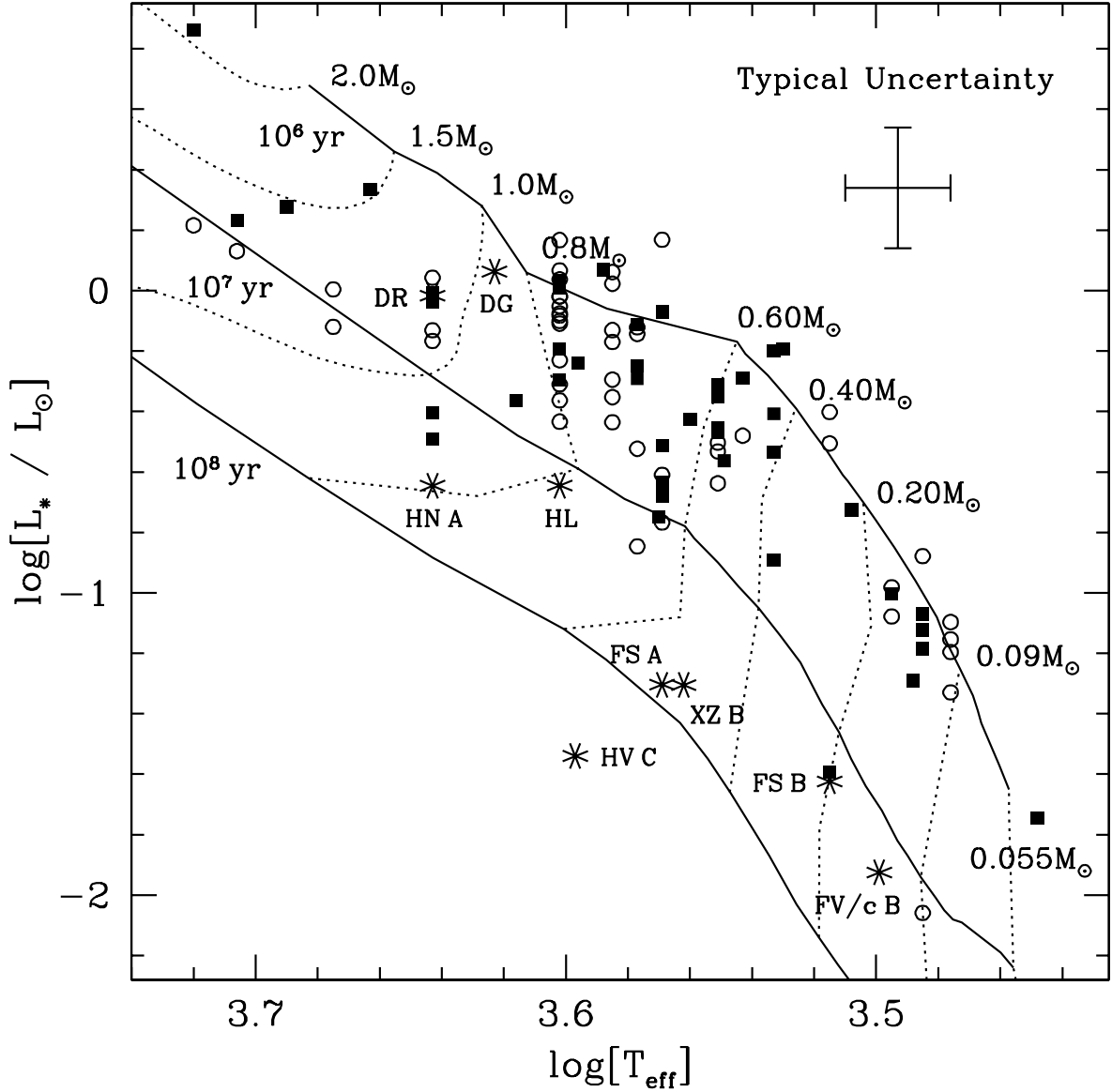}
\caption{
The components of the multiple star systems (\textit{squares}) and single
stars (\textit{circles}) are plotted on an H-R diagram along with adopted
PMS evolutionary models of Baraffe et al. (1998; M $< 1.0$ M$_\odot$) and
Palla \& Stahler (1998; M $> 1.0$ M$_\odot$).  Since the adopted method for
deriving luminosities and temperatures is inadequate for high accretion
stars (\S 3.1.1), these objects are distinguished as \textit{asterisks} and
labeled. \label{fig_bpsbin}}
\end{figure}

\begin{figure}
\epsscale{1.0}
\plotone{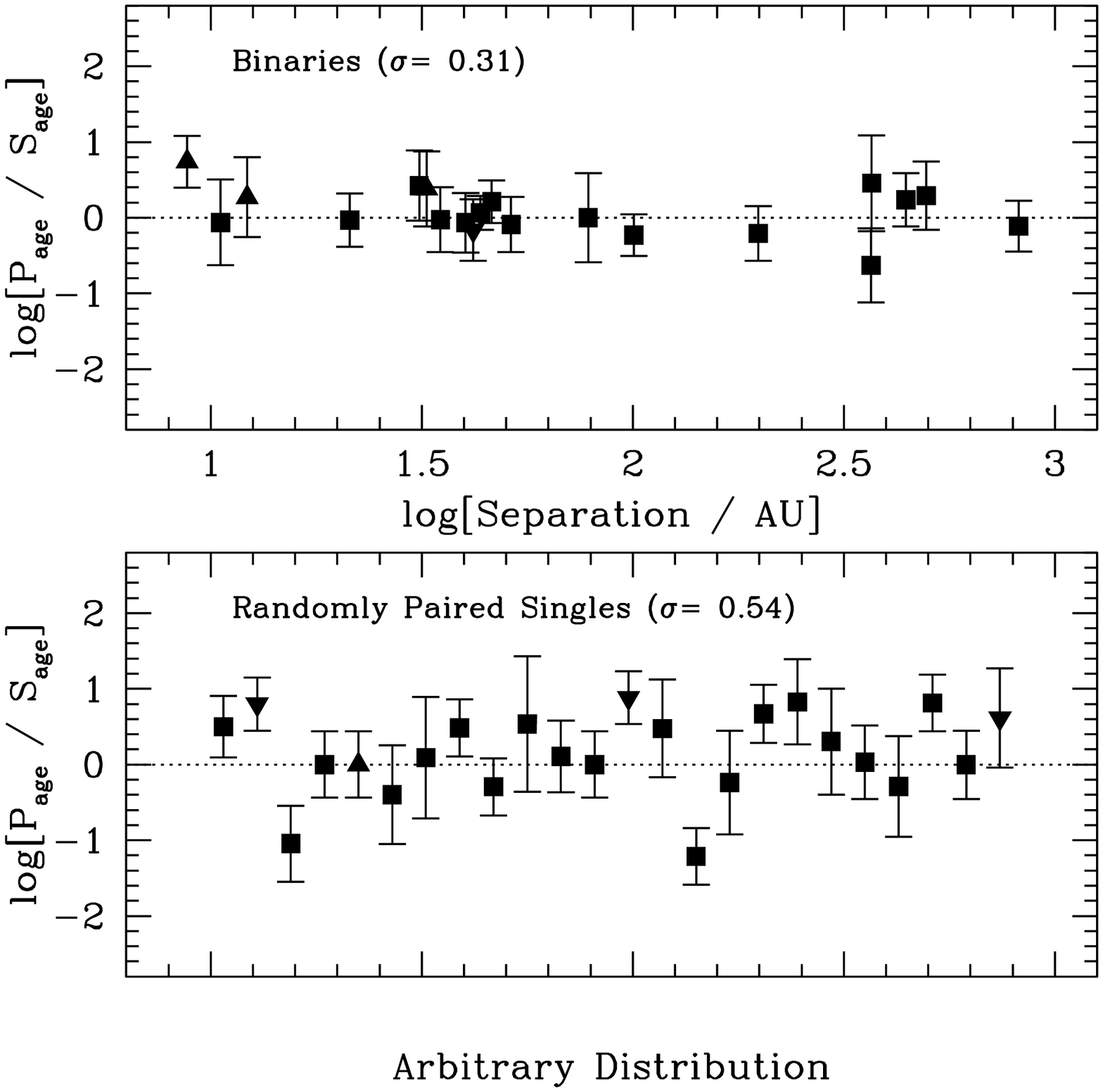}
\caption{The relative ages of binary star components are plotted versus
their projected separations.  The \textit{triangles} represent systems with
either primary age upper limits (\textit{downward pointing}) or secondary
age upper limits (\textit{upward pointing}).  There is no statistically
significant difference in age between primaries and secondaries over all
separation ranges.  The bottom panel shows the relative ages for a sample
of randomly paired single stars, plotted with arbitrary separations.  The
components of binary stars are generally more coeval than single stars
within the same star forming region. \label{fig_agesep}}
\end{figure} 

\begin{figure}
\epsscale{1.0}
\plotone{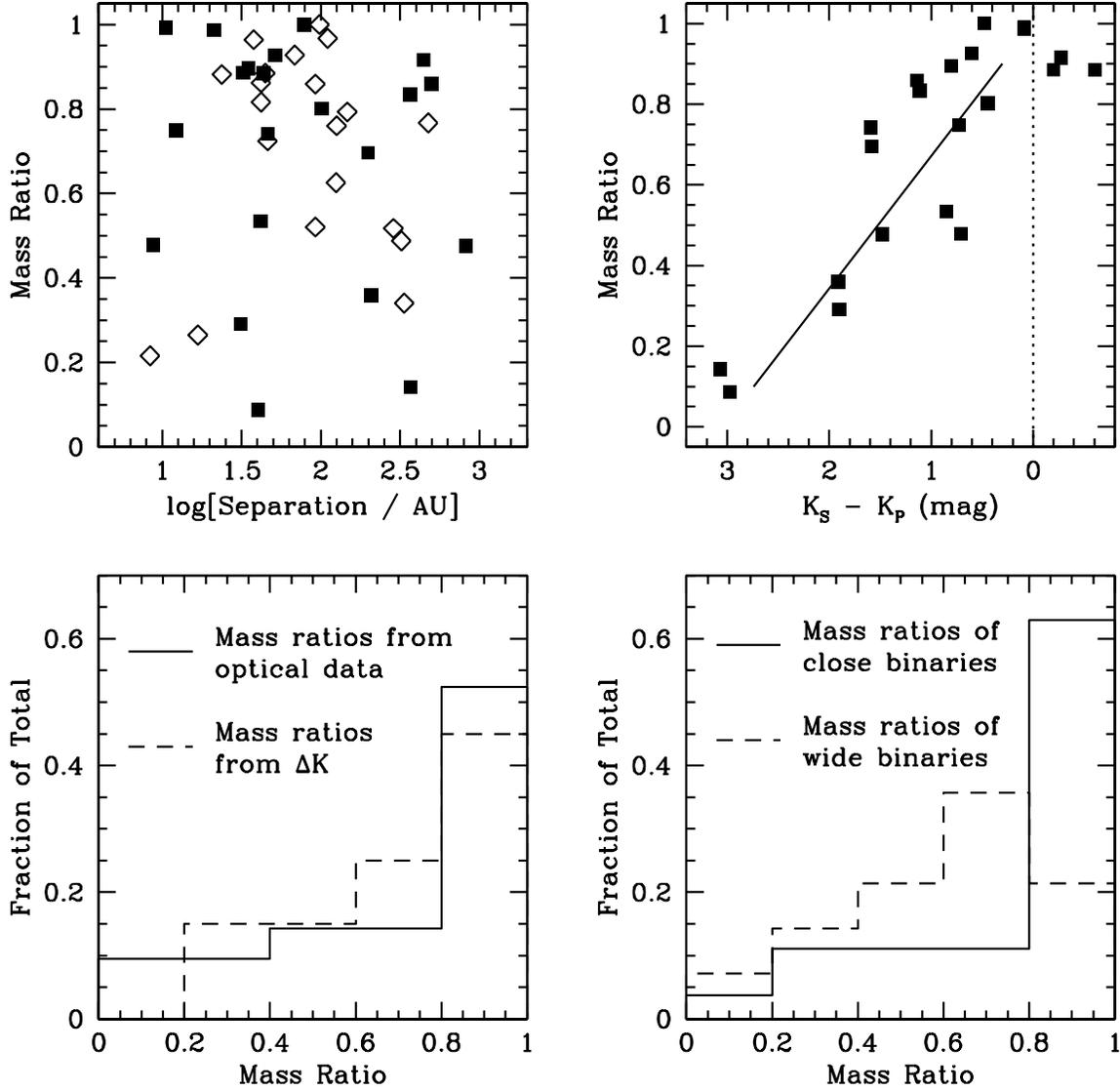}
\caption{In the top left panel, the mass ratios
(m$_\textrm{s}$/m$_\textrm{p}$) of pairs are plotted versus the their
projected separations.  \textit{Solid squares} represent mass ratios
computed from
optically resolved measurements (Table \ref{tab_bin_stel}).  \textit{Open
diamonds} represent mass ratios computed from $K$ magnitude differences,
following the linear relation shown in the top right panel
(m$_\textrm{s}$/m$_\textrm{p} = 1 - 0.328\Delta K$).  The bottom left panel
shows that mass ratios computed from optical measurements (\textit{solid
histogram}) and from $K$ magnitude differences (\textit{dashed histogram})
are similar and are peaked toward unity.  Combining mass ratios computed
from both methods, the bottom right panel shows that close pairs
(\textit{solid histogram}) have a higher fraction of high mass ratio
binaries than wide pairs (\textit{dashed histogram}). 
\label{fig_hst6_mass}}
\end{figure}

\begin{figure}
\epsscale{1.0}
\plotone{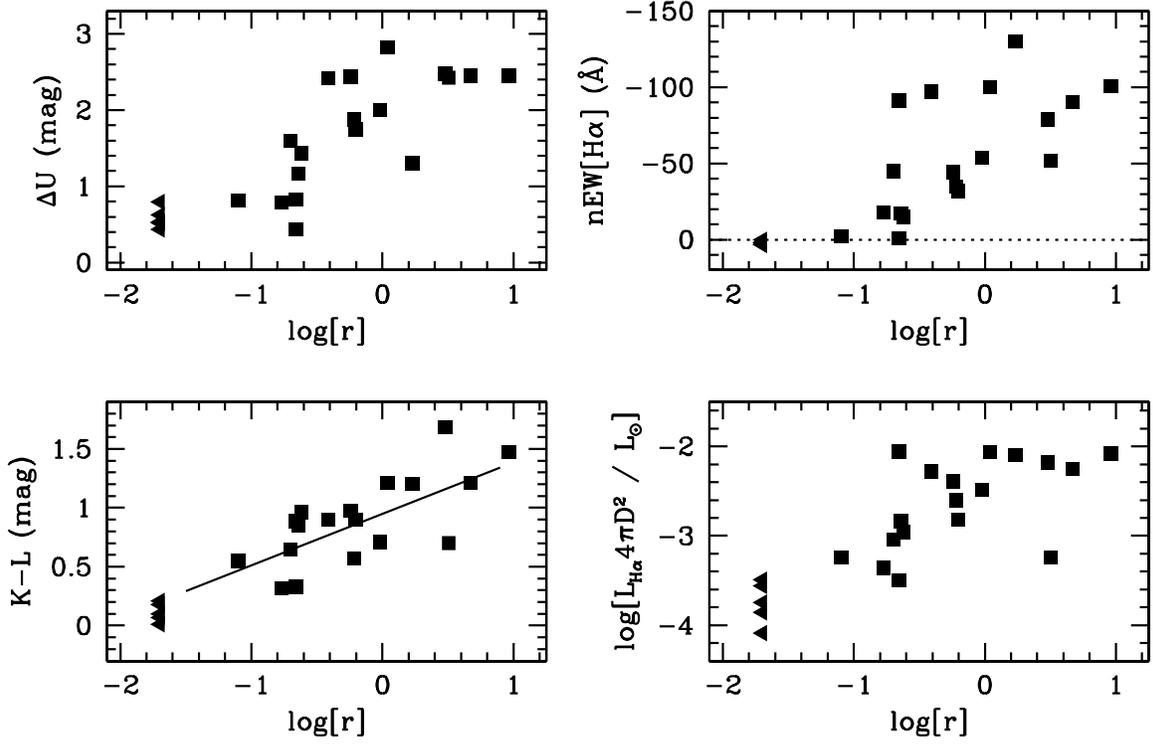}
\caption{The $\Delta U$s, $K-L$ colors, nEW(H$\alpha$)s, and H$\alpha$
luminosities (\S 3.2.1) of single T Tauri stars are compared to their levels
of optical excess emission (r = $F_{ex}/F_{phot}$) measured in high
resolution optical spectra \citep{heg95}.  \textit{Triangles} represent
stars with optical excess upper limits.  All diagnostics are at least
modestly correlated with the level of optical excess.  A linear least
squares fit to the $K-L$ versus log[r] relation is also shown (log[r] =
2.199($K-L$) - 2.148). \label{fig_duklha} }
\end{figure}
\clearpage

\begin{figure}
\epsscale{1.0}
\plotone{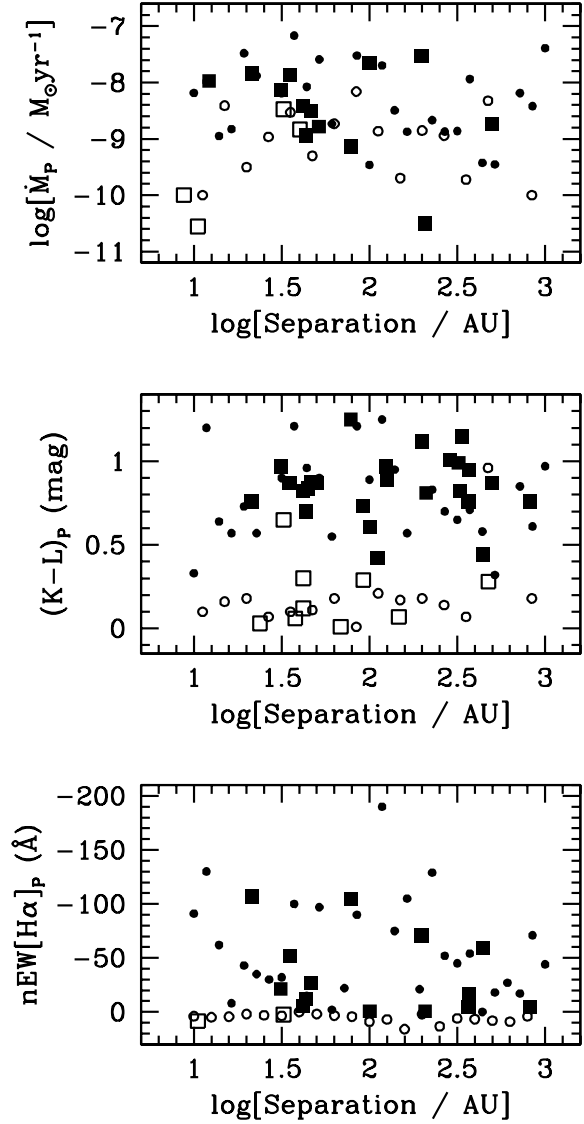}
\caption{
The mass accretion rates, the $K-L$ colors, and the nEW[H$\alpha$]s
of the CTTS primaries (\textit{large filled squares}) and WTTS primaries
(\textit{large open squares}) are plotted versus the projected separations
of their companions.  The corresponding accretion diagnostics of single
CTTSs (\textit{small filled circles}) and WTTSs (\textit{small open
circles}) are also plotted with an arbitrary distribution of separations
for comparison.  The accretion signatures of primaries and singles are
similar. \label{fig_pacc3} }
\end{figure}
\clearpage

\begin{figure}
\epsscale{1.0}
\plotone{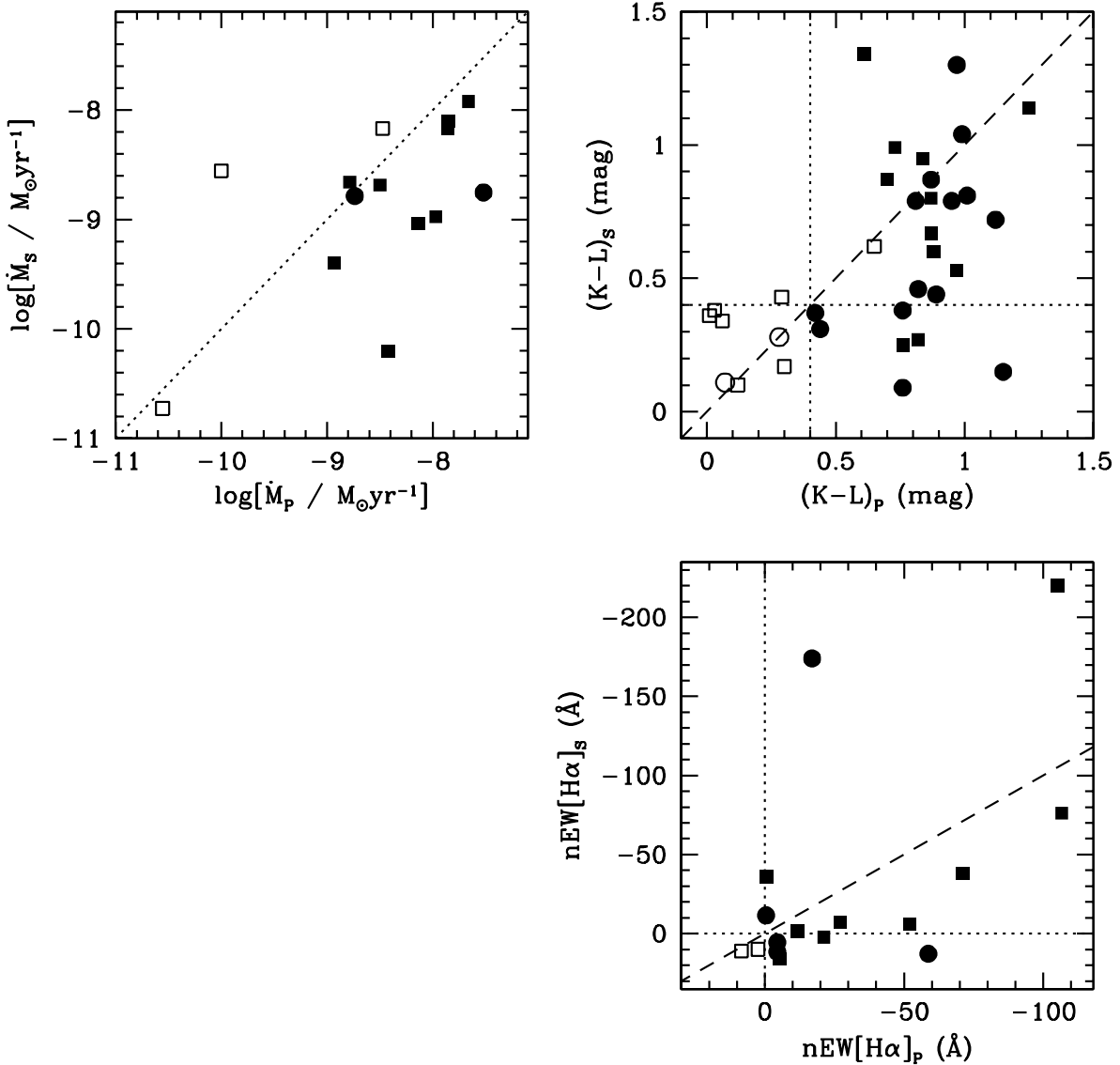}
\caption{
The mass accretion rates, the $K-L$ colors, and the nEW[H$\alpha$]s
of secondary stars are compared to primary stars.  CTTS systems
(\textit{filled symbols}) are distinguished from WTTS systems (\textit{open
symbols}) and close pairs (10-100 AU; \textit{squares}) are distinguished
from wide pairs (100-1000 AU; \textit{circles}).  $K-L$ colors greater than
0.4 magnitudes and nEW[H$\alpha$]s greater than zero imply the existence of
a circumstellar disk.  Primary stars appear to have, on average, accretion
signatures that are comparable to or larger than those of their companions.
Although several systems appear to harbor a circumprimary disk and no
circumsecondary disk, systems with only a circumsecondary disk are much
more rare. \label{fig_psacc3}} 
\end{figure} 
\clearpage

\begin{figure}
\epsscale{1.0}
\plotone{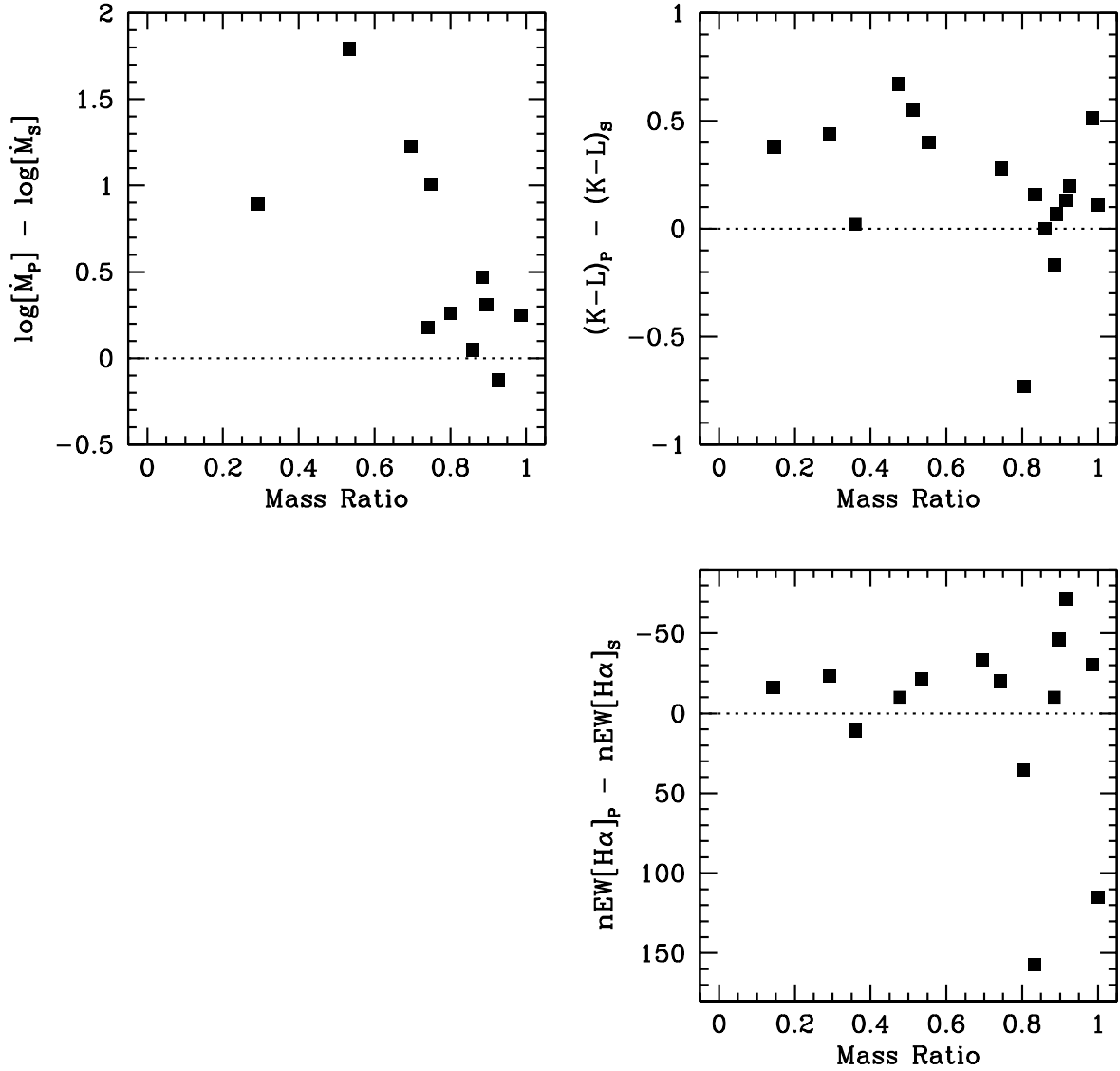}
\caption{The differences in the mass accretion rates, the $K-L$ colors, and
the nEW[H$\alpha$]s of primaries and secondaries are plotted versus the
binary mass ratios (m$_\textrm{s}$/m$_\textrm{p}$).  For pairs with
comparable masses (m$_\textrm{s}$/m$_\textrm{p}$ $>$ 0.8), the primaries
and secondaries have accretion signatures of comparable strength.  For
systems with small mass ratios (m$_\textrm{s}$/m$_\textrm{p}$ $<$ 0.8), the
primaries generally have the dominant accretion
signatures. \label{fig_qacc3} } 
\end{figure} 
\clearpage

\begin{figure}
\epsscale{1.0}
\plotone{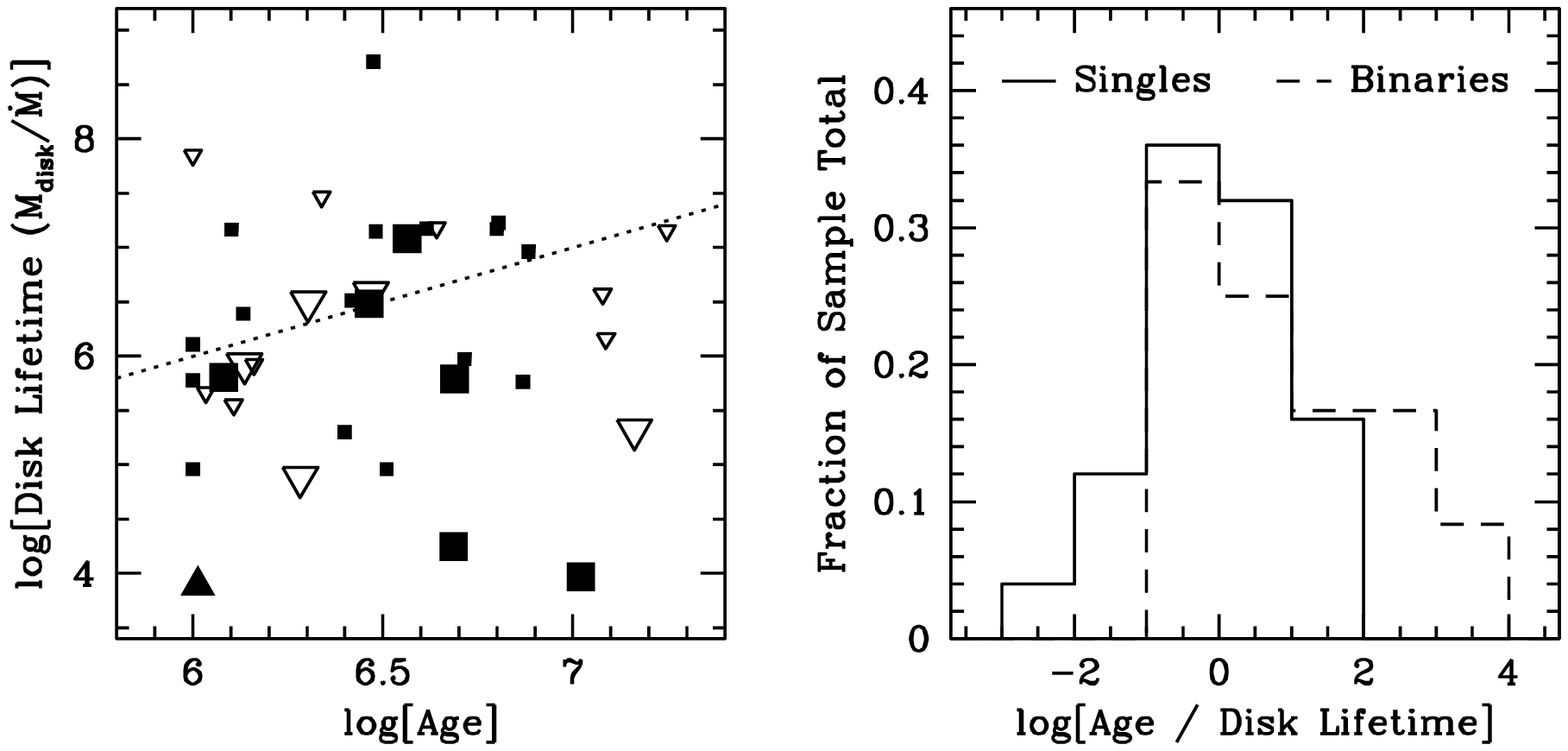}
\caption{In the left panel, the disk lifetimes (M$_\textrm{\scriptsize
disk}$/$\dot{\textrm M}$) of single (\textit{small symbols}) and binary
(\textit{large symbols}) T Tauri stars are compared to their stellar ages.
\textit{Solid squares} are calculated from disk mass estimates,
\textit{solid triangles} are calculated from disk mass lower limits, and
\textit{open triangles} are calculated from disk mass upper limits
\citep{ob95, beckwith90, dutrey96, gds99}.  The \textit{dotted line}
represents disk lifetimes equal to stellar ages.  In the right panel,
the distributions of log[Age/Disk Lifetime] for singles and binaries are
shown.  Single stars have, on average, disk lifetimes comparable to their
ages, and binary stars have, on average, disk lifetimes that are roughly
1/10 of their ages, but the distributions of log[Age/Disk Lifetime] for
singles and binaries are only marginally different.
\label{fig_disklife} }
\end{figure} 

\begin{figure}
\epsscale{1.0}
\plotone{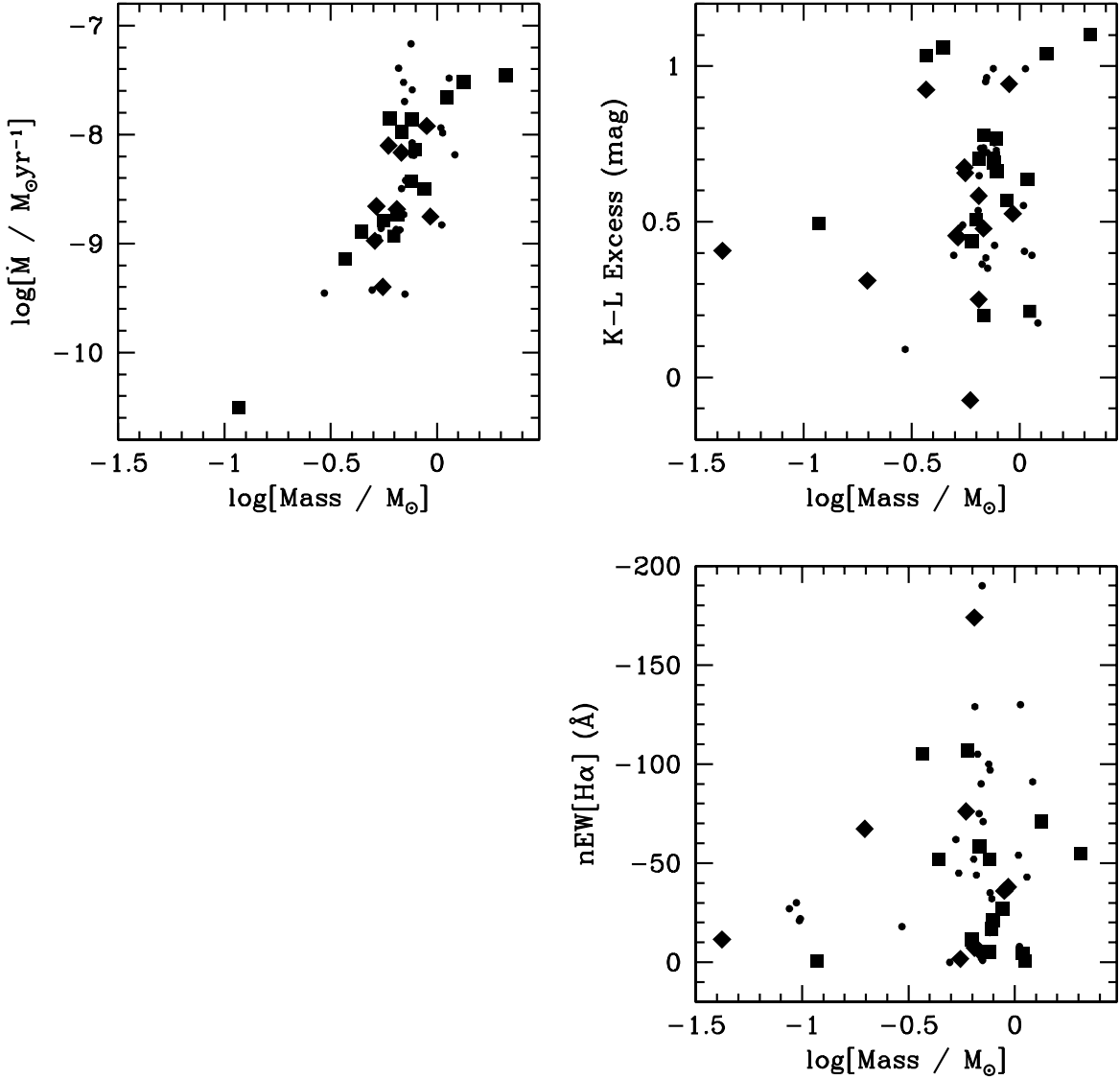}
\caption{The mass accretion rates, the $K-L$ color excesses ($(K-L)_{obs} -
(K-L)_{phot}$), and the nEW[H$\alpha$]s of CTTSs are plotted versus their
stellar mass.  Primaries (\textit{large squares}) are distinguished
from secondaries (\textit{large diamonds}) and from single stars
(\textit{small circles}).  The distribution of mass accretion rates shows a
general trend of decreasing mass accretion rates with decreasing stellar
masses.  Although the $K-L$ colors and nEW[H$\alpha$]s of the lowest mass
stars are below the mean of higher mass stars, these trends are suggestive,
but not statistically significant.
\label{fig_macc3} }
\end{figure} 

\clearpage

\begin{deluxetable}{llccll}
\tablecaption{Summary of Sample and Measurements \label{tab_sample}}
\tablewidth{0pt}
\tablehead{
\colhead{Source\tablenotemark{a}} & 
\colhead{Pair\tablenotemark{b}} &
\colhead{Sep} &
\colhead{Type\tablenotemark{c}} &
\colhead{Optical\tablenotemark{d}} &
\colhead{NIR\tablenotemark{d}} }
\startdata
\multicolumn{6}{c}{\underline{Close Binary Stars}} \\
V773 Tau& A$^\dag$-B	& 0\farcs07& W	&\textit{UBVRI}	& \textit{K} \\
V410 Tau& A-B	& 0\farcs07	& W	&\textit{\lbrack UBVRI\rbrack}	& \textit{K} \\
FW Tau	& A-B 	& 0\farcs08	& W	&UBVRI,H$\alpha$& \textit{K} \\
DF Tau	& A-B	& 0\farcs09	& C	&\textit{UBVRI}	& K \\
(RW Aur)& (Ba-Bb)& (0\farcs09)	& C	&\textit{\lbrack UBVRI\rbrack}	& \textit{K} \\
(FY Tau)& (A-B)	& ($> 0\farcs15$)& C 	&\lbrack UBVRI\rbrack		& \textit{K} \\
FO Tau	& A-B	& 0\farcs15	& C	&UBVRI,H$\alpha$& KL \\
\enddata
\tablenotetext{a}{\,RW Aur Ba-Bb and FY Tau A-B are listed in parentheses
because we have concluded they are not binary stars (\S 2.3.2).}
\tablenotetext{b}{\,A daggar ($^\dag$) indicates an unresolved companion
star from 
\citet[][V773 Tau]{welty95},
\citet[][RW Aur]{gahm99}, 
\citet[][J1-4872 B]{gsw01}, 
\citet[][Haro 6-37, J1-4872 A, UX Tau B]{duchene99a}, 
\citet[][UZ Tau A]{mathieu00},
\citet[][HV Tau, V807 Tau]{simon95} and
\citet[][T Tau]{koresko00}.}
\tablenotetext{c}{\,T Tauri type of the system (C=CTTS and W=WTTS).}
\tablenotetext{d}{\,Filters listed in italics denote measurements obtained
from the literature (\S 2.3.1 \& \S 2.3.3).  Filters listed in brackets
denote measurements in which the targeted companion was not resolved or
detected.}
\end{deluxetable}

\begin{landscape}
\begin{deluxetable}{llllllllll}
\tabletypesize{\scriptsize}
\tablecaption{Near Infrared Measurements of TTSs \label{tab_nir}}
\tablewidth{0pt}
\tablehead{
\colhead{} &
\colhead{} &
\colhead{Obs} &
\colhead{Flux} &
\colhead{System} &
\colhead{Primary} &
\colhead{Secondary} &
\colhead{Sep} &
\colhead{PA} &
\colhead{} \\
\colhead{Source} & 
\colhead{Flt} & 
\colhead{Date} & 
\colhead{Ratio\tablenotemark{a}}	& 
\colhead{(mag)} & 
\colhead{(mag)} & 
\colhead{(mag)} & 
\colhead{(arcsec)} &
\colhead{(degrees)} &
\colhead{Ref} }
\startdata
DF Tau A-B& K & 1996 Dec 6 & $1.62 \pm 0.17$ & \nodata & \nodata & 
\nodata & $0.0896 \pm 0.0079$	& $290.2 \pm 1.2$	& 1 \\

FO Tau A-B& K & 1996 Dec 5 & $1.000 \pm 0.010$ & \nodata & \nodata &
\nodata & $0.1499 \pm 0.0069$	& $193.7 \pm 1.0$	& 1 \\
& & & & & $8.89 \pm 0.14^*$ & $8.98 \pm 0.15^*$ & & & 1,2,3,4 \\
	& L & 1996 Dec 6 & $1.74 \pm 0.49$ & \nodata & \nodata &
\nodata & $0.160 \pm 0.010$	& $188.4 \pm 5.9$	&  1 \\
& & & & & $8.13 \pm 0.12^*$ & $8.73 \pm 0.20^*$ & & & 1,4 \\

V928 Tau A-B& K & 1997 Dec 9 & $1.055 \pm 0.037$ & \nodata & \nodata &
\nodata & $0.1967 \pm 0.0037$ & $117.2 \pm 1.2$ & 1 \\
& & & & & $8.64 \pm 0.09^*$ & $9.00 \pm 0.12^*$ & & & 1,2,3,4 \\
	& L & 1997 Dec 8 & $1.009 \pm 0.002$ & $7.95 \pm 0.05$ & 
$8.70 \pm 0.05$ & $8.71 \pm 0.05$ & $0.1851 \pm 0.0035$ & $118.7 \pm 1.8$
& 1 \\
\enddata
\tablecomments{The $K$ and $L$ magnitudes used in the analysis (\S 3 \& \S
4) are marked with asterisks and are the combined values from the listed
references.}
\tablenotetext{a}{\,Flux ratios are calculated as A/B, where A and B are
defined in \S 2.1.  This is equivalent to primary/secondary for all pairs
except LkHa 332/G1.}
\tablenotetext{b}{\,The brighter component of LkHa 332/G1, listed under the
column labeled Primary, is actually the less massive component, but is
treated as component A here to avoid confusion with previous measurements
(see \S 2.1).}
\tablerefs{
(1) this work;
(2) \citet{gnm93};
(3) \citet{leinert93};
(4) \citet{kh95};
(5) \citet{chen90};
(6) \citet{simon92};
(7) \citet{leinert91};
(8) \citet{ghez95b};
(9) \citet{roddier96};
(10) \citet{gws97};
(11) \citet{haas90};
(12) \citet{ghez91};
(13) \citet{lh89};
(14) \citet{menard93};
(15) \citet{hss94};
(16) \citet{mz91};
(17) \citet{wl98}
}
\end{deluxetable}
\end{landscape}

\begin{landscape}
\begin{deluxetable}{llllrllll}
\tabletypesize{\scriptsize}
\tablecaption{HST/WFPC2 Observations of Binary T Tauri Stars
\label{tab_wfpc2_bin} }
\tablewidth{0pt}
\tablehead{
\colhead{Source} &
\colhead{HST} &
\colhead{} &
\colhead{System} &
\colhead{Trans-} &
\colhead{Primary} &
\colhead{Secondary} &
\colhead{Mean Sep} &
\colhead{Mean PA} \\
\colhead{(Obs Date)} & 
\colhead{Flt} & 
\colhead{Flux Ratio\tablenotemark{a}} & 
\colhead{(mag)}	& 
\colhead{formed to} &
\colhead{(mag -$or$- \AA)} & 
\colhead{(mag -$or$- \AA)} & 
\colhead{(arcsec)} & 
\colhead{(degrees) } }
\startdata
V773 Tau A-B	& F336W & $5.7 \pm 4.1$ & $13.36 \pm 0.03$ & $U$
	& $13.60 \pm 0.12$ & $15.49 \pm 0.67$ \\
(1994 Oct 29)	& F439W & $13 \pm 11$ & $12.38 \pm 0.02$ & $B$
	& $12.31 \pm 0.07$ & $15.10 \pm 0.85$ \\
		& F555W & $4.9 \pm 2.0$ & $10.84 \pm 0.03$ & $V$
	& $11.03 \pm 0.08$ & $12.76 \pm 0.37$ \\
		& F675W & $3.5 \pm 1.2$ & $9.76 \pm 0.01$ & $R_c$
	& $10.13 \pm 0.08$ & $11.49 \pm 0.29$ \\
		& F814W & $3.0 \pm 1.5$ & $9.03 \pm 0.03$ & $I_c$
	& $9.34 \pm 0.14$ & $10.53 \pm 0.41$ 
	& $0.0628 \pm 0.0024$ & $321.9 \pm 2.7$ \\
\\

V410 A-B	& F336W & $> 207$ & $12.84 \pm 0.01$ & $U$
	& $12.91 \pm 0.02$ & $> 16.2$ \\
(1994 July 31)	& F439W & $> 780$ & $12.05 \pm 0.02$ & $B$
	& $11.88 \pm 0.02$ & $> 15.6$ \\
\enddata
\tablenotetext{a}{\,Flux ratios are calculated as A/B, where A and B are
	defined in \S 2.1.  This is equivalent to primary/secondary for all
	pairs except LkHa 332/G1.}
\tablenotetext{c}{\,The optical measurements of FS Tau may be biased because
	of its extended emission.}
\tablenotetext{c}{\,The brighter component of LkHa 332/G1, listed under
	the column labeled Primary, is actually the less massive component,
	but is treated as component A here to avoid confusion with
	previous measurements (see \S 2.1).}
\end{deluxetable}
\end{landscape}

\begin{deluxetable}{llrlll}
\tabletypesize{\scriptsize}
\tablecaption{HST/WFPC2 Observations of FS Tau \label{tab_wfpc2_fs}}
\tablewidth{0pt}
\tablehead{
\colhead{HST} &
\colhead{Total\tablenotemark{a}} &
\colhead{Trans-} &
\colhead{Total} &
\colhead{Binary\tablenotemark{b}} &
\colhead{Nebula\tablenotemark{c}} \\
\colhead{Flt} & 
\colhead{(mag)}	& 
\colhead{formed to} & 
\colhead{(mag -$or$- \AA)} &
\colhead{(mag -$or$- \AA)} &
\colhead{(mag -$or$- \AA)} }
\startdata
F336W	& $17.15 \pm 0.25$ & $U$ & $17.21 \pm 0.25$ & $18.81 \pm 0.15$ &
	$17.49 \pm 0.32$ \\
F439W	& $17.42 \pm 0.06$ & $B$ & $17.18 \pm 0.07$ & $19.02 \pm 0.05$ &
	$17.40 \pm 0.08$ \\
F555W	& $15.81 \pm 0.08$ & $V$ & $15.80 \pm 0.08$ & $17.53 \pm 0.07$ &
	$16.04 \pm 0.11$ \\
F675W	& $14.36 \pm 0.09$ & $R_c$ & $14.48 \pm 0.09$ & $15.99 \pm 0.09$ &
	$14.80 \pm 0.12$ \\
F814W	& $13.11 \pm 0.02$ & $I_c$ & $13.10 \pm 0.02$ & $14.34 \pm 0.05$ &
	$13.52 \pm 0.03$ \\
F656N	& $12.51 \pm 0.03$ & EW[H$\alpha$] & $-37.8 \pm 5.7$ & $-72.1 \pm 3.5$ &
	$-40.8 \pm 8.0$ \\
\enddata
\tablenotetext{a}{\,A 4\farcs0 radius is used to determine the Total
magnitude (\S 2.3.3).}
\tablenotetext{b}{\,The Binary magnitudes are calculated from the
sum of the component flux densities (Table \ref{tab_wfpc2_bin}).}
\tablenotetext{c}{\,The Nebula magnitudes are calculated from the
differences in the Total and Binary flux densities.}

\end{deluxetable}

\begin{landscape}
\begin{deluxetable}{llllclrccccccc}
\tabletypesize{\tiny}
\tablecaption{Stellar and Circumstellar Properties of Optically Resolved
Systems \label{tab_bin_stel} }
\tablewidth{0pt}
\tablehead{
\colhead{} &
\colhead{} &
\colhead{} &
\colhead{} &
\colhead{SpT} &
\colhead{$A_\textrm{v}$} &
\colhead{} &
\colhead{Mass} &
\colhead{} &
\colhead{EW[H$\alpha$]} &
\colhead{$\Delta U$} &
\colhead{} &
\colhead{$K-L$} &
\colhead{Tauri}
\\
\colhead{Source} & 
\colhead{} &
\colhead{SpT\tablenotemark{a,b}} &
\colhead{log[T]\tablenotemark{b}} &
\colhead{Ref} &
\colhead{(mag)} & 
\colhead{log[$\frac{\textrm L}{\textrm L_\odot}$]} &
\colhead{(M$_\odot$)} &
\colhead{log[$\frac{\textrm{Age}}{\textrm{yr}}$]} &
\colhead{(\AA)} &
\colhead{(mag)} &
\colhead{log[$\frac{\dot{\textrm M}}{\textrm M_\odot\textrm{yr}^{-1}}$]} &
\colhead{(mag)} &
\colhead{Type\tablenotemark{c}} }
\startdata
V773 Tau& A	& K2	& 3.690 			& 1 & $1.39 \pm 0.17$ & $0.276 \pm 0.089$ & 
	$1.46 \pm 0.10$	& $6.74 \pm 0.17$ &\nodata	& -0.09 &$<-10$ 
	&\nodata & W::\\
	& C	& \textit{M0}	& \textit{3.588}	& 2 & $1.39$	& $0.070 \pm 0.18$ & 
	$0.70 \pm 0.07$	& $< 6.0 \pm 0.3$ &\nodata	& 0.63 &$<-8.6$ 
	&\nodata & W::\\
FW Tau	& A	& M5.5$\pm 0.5$ & 3.485			& 3 & $0.00 \pm 0.83$ & $-1.07 \pm 0.34$ & 
	$0.12 \pm 0.04$	& $6.09 \pm 0.41$ & -11.6	& 0.86 &$<-10.6$
	&\nodata & W\\
	& B	& \textit{M5.5}	& \textit{3.485}	& 2 & $0.00$	& $-1.12 \pm 0.34$ & 
	$0.12 \pm 0.08$	& $6.15 \pm 0.39$ & -9.0	& 0.75 &$<-10.7$
	&\nodata & W\\
	& C	&\nodata&\nodata&	&\nodata	& \nodata
	&\nodata	&\nodata	&$< -454$	&\nodata &\nodata	
	&\nodata & C\\
DF Tau	& A	& M0.5	& 3.577 			& 4 & $0.15 \pm 0.80$	& $-0.25 \pm 0.32$ & 
	$0.68 \pm 0.07$	& $6.27 \pm 0.53$ &\nodata	& 2.43 & -8.0		
	&\nodata& C::\\
	& B	& \textit{M3}	& \textit{3.533}	& 2 & $0.15$	& $-0.20 \pm 0.32$ & 
	$0.51 \pm 0.12$	& $< 6.0 \pm 0.3$&\nodata	& 0.85 & -9.0	
	&\nodata& C::\\
FO Tau	& A	& M2	& 3.551 			& 4 & $3.03 \pm 0.51$	& $-0.32 \pm 0.21$ &
	$0.60 \pm 0.09$	& $6.27 \pm 0.27$	& -117	& 3.14 & -7.9		
	& 0.76	& C\\
\enddata
\tablecomments{The parameters derived for high accretion stars are listed
in parentheses and are subject to large uncertainties (\S 3.1.1).}
\tablenotetext{a}{\,Spectral type uncertainties are one spectral subclass
unless otherwise noted.}
\tablenotetext{b}{\,Spectral types and temperatures listed in italics are
derived from photometry (\S 3.1).}
\tablenotetext{c}{\,High accretion stars (C*) are identified using the $K-L$
colors while CTTSs (C) and WTTSs (W) are distinguished based on the
EW[H$\alpha$]s or, if the EW[H$\alpha$] is not available, $K-L$ colors
(listed with a colon) or $U$-band excesses (listed with a double colon; \S
3.2.2).}
\tablenotetext{d}{\,The veiled optical spectrum of XZ Tau B suggests that its
spectral type is cooler than M1 (\S 2.4.2), which is consistent with the
photometric value listed.}
\tablenotetext{e}{\,Multiple EW[H$\alpha$] measurements are from WFPC2 (\S
2.2) and FOS (\S 2.3), respectively; the average value is used in the
analysis.}
\tablerefs{
(1) \citet{welty95};
(2) photometry in \S 3.1;
(3) \citet{briceno98};
(4) \citet{ck79};
(5) \citet{martin94};
(6) \citet{hss94};
(7) \citet{white99};
(8) \citet{rs96};
(9) spectra in \S 2.3; 
(10) \citet{bb90};
(11) \citet{mg82};
(12) \citet{duchene99b};
(13) \citet{luhman99};
(14) \citet{bm95}.
}
\end{deluxetable}
\end{landscape}

\begin{deluxetable}{lccccccl}
\tabletypesize{\scriptsize}
\tablecaption{HST Filter Transformations \label{tab_filt}}
\tablewidth{0pt}
\tablehead{
\colhead{} &
\colhead{} &
\colhead{} &
\colhead{} &
\colhead{} &
\colhead{} &
\colhead{} &
\colhead{}
\\
\colhead{SpT} &
\colhead{$U$-F336W} &
\colhead{$B$-F439W} &
\colhead{$V$-F555W} &
\colhead{$R_c$-F675W} &
\colhead{$I_c$-F814W} &
\colhead{F656N-F675W} &
\colhead{Star ID\tablenotemark{a}} }
\startdata
G8	& 0.07	& -0.14	& -0.01	& 0.10	& 0.00	& -0.24 & 53, 54 \\
K0	& 0.07	& -0.14	& -0.01	& 0.10	& 0.00	& -0.24 & \\
K1	& 0.07	& -0.15	& -0.01	& 0.10	& 0.00	& -0.24 & \\
K2	& 0.07	& -0.15	& -0.01	& 0.10	& 0.00	& -0.24 & \\
K3	& 0.07	& -0.15	& -0.01	& 0.10	& 0.00	& -0.24 & 56, 57 \\
K4	& 0.07	& -0.17	& -0.01	& 0.11	& 0.00	& -0.25 & \\
K5	& 0.08	& -0.19	& -0.01	& 0.11	& -0.01	& -0.27 & \\
K6	& 0.08	& -0.21	& -0.01	& 0.12	& -0.01	& -0.29 & \\
K7	& 0.08	& -0.23	& -0.01	& 0.13	& -0.01 & -0.31 & 63, 64 \\
M0	& 0.07	& -0.24	& -0.01	& 0.13	& -0.01	& -0.30 & 61, 65, 66 \\
M1	& 0.06	& -0.24	& -0.01	& 0.13	& -0.01 & -0.31 & \\
M2	& 0.06	& -0.25	& -0.02	& 0.13	& -0.01 & -0.33 & 67 \\
M3	& 0.06	& -0.27	& -0.02	& 0.13	& -0.01 & -0.34 & \\
M4	& 0.05	& -0.29	& -0.03	& 0.12	& -0.01 & -0.35 & 68 \\
M5	& 0.17	& -0.35	& -0.03	& 0.11	& 0.00	& -0.37 & \\
M6	& 0.28	& -0.40	& -0.03	& 0.11	& 0.00	& -0.38 & 69 \\
M7	& 0.24	& -0.36	& -0.04	& 0.11	& -0.02	& -0.43 & \\
M8	& 0.20	& -0.31	& -0.05	& 0.11	& -0.04	& -0.47 & 70 \\
\enddata
\tablenotetext{a}{\,The ID number of the stars from \citet{gs83} used to
establish the transformations.}
\end{deluxetable}

\begin{deluxetable}{llllcclcccc}
\tabletypesize{\tiny}
\tablecaption{Single T Tauri Stars in Taurus-Auriga \label{tab_sing}}
\tablewidth{0pt}
\tablehead{
\colhead{} &
\colhead{} &
\colhead{EW[H$\alpha$]} &
\colhead{} &
\colhead{$A_\textrm{v}$} &
\colhead{} &
\colhead{Mass} &
\colhead{} &
\colhead{$\Delta U$} &
\colhead{} &
\colhead{$K-L$}
\\
\colhead{Source} & 
\colhead{SpT\tablenotemark{a}} &
\colhead{(\AA)} &
\colhead{Ref} &
\colhead{(mag)} & 
\colhead{log[$\frac{\textrm L}{\textrm L_\odot}$]} & 
\colhead{(M$_\odot$)} &
\colhead{log[$\frac{\textrm{Age}}{\textrm{yr}}$]} &
\colhead{(mag)} &
\colhead{log[$\frac{\dot{\textrm M}}{\textrm M_\odot\textrm{yr}^{-1}}$]} &
\colhead{(mag)} }
\startdata
\\
\multicolumn{11}{c}{\underline{High Accretion T Tauri Stars}} \\
DR Tau		& K5	& -106	& 1,2		&(0.51)	&(-0.017)& (1.11)&
(6.60) & (2.45) & (-7.50) & 1.47\\
DG Tau		& K6	& -84	& 1,2		&(1.41)	&(0.062)& (0.88)&
(6.12) & (2.48) & (-7.34) & 1.68\\
HL Tau		& K7	& -55	& 3,2		&(1.82)	&(-0.646)& (0.78)&
(7.15) & (1.79) & (-8.83) & 1.76\\
\\
\multicolumn{11}{c}{\underline{Classical T Tauri Stars}} \\
GM Aur		& K3	& -96	& 3,2		&1.21	& 0.004	& 1.22	&
6.87 & 0.83 & -8.18 & 0.33\\
\enddata
\tablecomments{The parameters derived for high accretion stars are
listed in parentheses and are subject to large uncertainties (\S
3.1.1).}
\tablenotetext{a}{\,Spectral type uncertainties are one spectral subclass.}
\tablerefs{
(1) \citet{hg97};
(2) \citet{kh95}; 
(3) \citet{hb88}; 
(4) \citet{hss94}; 
(5) this work;
(6) \citet{ss94};
(7) \citet{briceno98};
(8) \citet{walter88}; 
(9) \citet{martin94};
(10) \citet{luhman98}}
\end{deluxetable}

\clearpage

\end{document}